\documentclass[preprint,review,11pt]{elsarticle}

\usepackage{lineno,hyperref}
\modulolinenumbers[5]

\journal{Combustion and Flame}

\usepackage{epstopdf}
\usepackage{graphicx}
\usepackage{color}
\usepackage{soul}

 \pdfoutput=1 
\usepackage{geometry}
\usepackage{setspace} 

\usepackage{amssymb}
\usepackage{amsmath}
\usepackage{multicol}
\usepackage{latexsym}
\usepackage{times}
\usepackage{subfigure}
\usepackage{multirow}
\usepackage{setspace}
\graphicspath{{Pictures/}}

\usepackage{longtable}
\usepackage[labelsep=period]{caption}
\biboptions{sort&compress}

\usepackage[table]{xcolor}

\usepackage{slashbox}
\usepackage{multirow}
\usepackage{arydshln}
\usepackage{amsmath,amssymb}

\usepackage{tikz}
\usetikzlibrary{calc,topaths}
\usepackage{notoccite}

\usepackage{setspace} 
\makeatletter
\def\ignorecitefornumbering#1{%
	\begingroup
	\@fileswfalse
	#1
	\endgroup
}
\makeatother

\begin{document}

\begin{frontmatter}



\title{\LARGE Contributions of flame thickening and local extinctions to burning rate of intensely turbulent premixed flames}

\author{{\large Sajjad Mohammadnejad$^{\mathrm{a}}$, Qiang An$^{\mathrm{a,b}}$, Patrizio Vena$^{\mathrm{b}}$, Sean Yun$^{\mathrm{b}}$, Sina Kheirkhah$^{\mathrm{a,*}}$\\[10pt]
		{\footnotesize \em $\mathrm{^a}$School of Engineering, University of British Columbia, Kelowna, Canada, V1V 1V7}\\[0pt]
		{\footnotesize \em $\mathrm{^b}$Gas Turbine Laboratory, Aerospace Research Centre, National Research Council, Ottawa, Canada, K1A 0R6}\\[-5pt]}}
\date{}


\cortext[cor1]{Corresponding author: sina.kheirkhah@ubc.ca}

\begin{abstract}

Influences of reaction zone thickening and local extinctions on the burning rate of extremely turbulent hydrogen-enriched methane-air flames are investigated using simultaneous planar laser-induced fluorescence of formaldehyde molecule and hydroxyl radical as well as separate stereoscopic particle image velocimetry techniques. The fuel-air equivalence ratio is set to 0.7, and the amount of hydrogen-enrichment varies from 0\% to 70\%. Mean bulk flow velocities ranging from 5 to 35~m/s with Reynolds and Karlovitz numbers of 18 to 2729 and 0.1 to 76.0, respectively, are examined. It is shown that, by increasing the turbulence intensity, the preheat and reaction zone thicknesses can increase to values that are, respectively, 6.3 and 4.9 of the corresponding laminar flames. Broadening of these zones for intensely turbulent hydrogen-enriched methane-air flames is shown experimentally for the first time. Broadening of the reaction zone suggests that the flamelet assumption used for development of the burning rate formulations may not hold. Thus, a new formulation, which does not utilize the flamelet assumption, is developed and used to calculate the burning rate of the tested flames. It is shown that, at small turbulence intensities, the burning rate values follow those of the local consumption speed, which is developed in the literature based on the flamelet assumption. However, at large turbulence intensities, the estimated burning rate features large values, and the ratio of this parameter to the local consumption speed is consistent with the ratio of the global and local consumption speeds reported in the literature. It is shown that the ratio of the normalized burning rate to the normalized local consumption speed is correlated with the broadening of reaction zone, suggesting that the disparity between the values of the burning rate and local consumption speed is linked to the reaction zone thickening. It is shown, although the flame thickening increase the burning rate, local extinctions decrease this parameter leading to the bending behavior reported in the literature. It is argued the amount of bending is positively and negatively related to the turbulence intensity and the integral length scale, respectively. Using this, a mathematical formulation that allows for estimation of the burning rate is developed. The predictions of the proposed formulation agree well with the experimentally estimated burning rate values.

\end{abstract}


\end{frontmatter}

\section{Introduction}
\label{Int}

Hydrogen-enrichment of natural gas (which mainly contains methane) will be used as a means to decarbonize several industries that use combustion equipment, such as land-based power generation gas turbine engines in the future~\cite{masri2020challenges}. While this is underway, our understanding related to the speed at which the fuel and air mixture is converted to combustion products (referred to as the turbulent burning velocity) at engine flow-relevant conditions is limited for both pure and hydrogen-enriched methane-air flames to our best of knowledge. The burning velocity has been estimated using the flamelet assumption for methane-air flames, see for example the review papers by Driscoll~\cite{driscoll2008turbulent,driscoll2020premixed}. This assumption suggests that the reaction zone of the premixed flames is relatively thin~\cite{driscoll2008turbulent,driscoll2020premixed,peters1988laminar,van2016state,gicquel2012large,duclos1993comparison,mohammadnejad2019internal}. Although several studies show this assumption holds for a wide range of turbulence conditions~\cite{skiba2018premixed,skiba2020experimental}, results of several other investigations~\cite{mohammadnejad2020thick,zhou2017thin,zhou2015distributed,zhou2015simultaneous,dunn2007new,dunn2009compositional,dunn2010finite,wang2018direct,wang2017direct,zhou2015visualization,wang2019structure} disagree and suggest the flamelet assumption may not hold. For these studies~\cite{mohammadnejad2020thick,zhou2017thin,zhou2015distributed,zhou2015simultaneous,dunn2007new,dunn2009compositional,dunn2010finite,wang2018direct,wang2017direct,zhou2015visualization,wang2019structure}, the turbulent burning velocity cannot be estimated using the formulations developed based on the flamelet assumption. The present investigation is motivated by development and assessment of a formulation that allows for calculation of the burning velocity for turbulent premixed flames whose internal structure do not necessarily follow the prediction of the flamelet assumption for both pure and hydrogen-enriched methane-air turbulent premixed flames. In the following, first, literature related to the effect of the turbulent flow on the flame structure, as well as the implication of this effect on the burning velocity of methane-air premixed flames is briefly reviewed. Then, background related to the effect of hydrogen-enrichment on the methane-air premixed flames internal structure and burning velocity is presented.


\subsection{Effect of turbulence on the flame internal structure and burning velocity}

The internal structure of premixed flames can be highly influenced by the Karlovitz number, $Ka=\tau_\mathrm{f}/\tau_{{\eta}_\mathrm{K}}$, where $\tau_\mathrm{f}$ and $\tau_{{\eta}_\mathrm{K}}$ are the flame and Kolmogorov time scales, respectively~\cite{peters2000turbulent,poinsot2005theoretical}. Flames with $Ka \lesssim 1$ usually feature both thin preheat and reaction zones, and the turbulent eddies only tend to wrinkle the flame surfaces~\cite{peters2000turbulent,bell2007numerical,buschmann1996measurement,filatyev2005measured}. For $Ka \gtrsim 1$, the results reported in the literature for the preheat zone thickness are controversial. On one hand, as predicted by Peters~\cite{peters2000turbulent}, for $1 \lesssim Ka \lesssim 100$, small eddies may penetrate into and broaden the preheat zone. This is confirmed through both experimental~\cite{osborne2016simultaneous,osborne2017relationship,skiba2018premixed,zhou2017thin,kariuki2015heat,kariuki2016heat,mansour1998investigation,mohammadnejad2020thick} and Direct Numerical Simulation (DNS) studies~\cite{poludnenko2010interaction,poludnenko2011interaction}. However, several studies~\cite{tamadonfar2015experimental,buschmann1996measurement,soika1998measurement,halter2008investigations,skiba2018premixed} suggest that the preheat zone may not broaden for $Ka \gtrsim 1$. The reason for this discrepancy is elaborated in~\cite{skiba2018premixed,driscoll2020premixed}. Driscoll et al.~\cite{driscoll2020premixed} and Skiba et al.~\cite{skiba2018premixed} argue that, eddies associated with $Ka \gtrsim 1$ may feature sufficient turbulent kinetic energy allowing them to penetrate and broaden the preheat zone. Skiba et al.~\cite{skiba2018premixed} and Driscoll et al.~\cite{driscoll2020premixed} suggested that, in addition to $Ka \gtrsim 1$, flames with $u'\Lambda/(S_\mathrm{L,0}\delta_\mathrm{L}) \gtrsim 180$  may feature broadened preheat zones, where $\Lambda$, $S_\mathrm{L,0}$, and $\delta_\mathrm{L}$ are the integral length scale, unstretched laminar flame speed, and laminar flame thickness, respectively~\cite{driscoll2020premixed,skiba2018premixed}.

Similar to the preheat zone, the results reported in the literature regarding the turbulent flames reaction zone thickness are also controversial. Experimental~\cite{skiba2018premixed} and DNS studies of~\cite{aspden2019towards,aspden2011lewis,aspden2011turbulence} show that the reaction zone remains thin upto $Ka \approx 550$ and 1000, respectively. In contrast to the studies of~\cite{skiba2018premixed,aspden2019towards,aspden2011lewis,aspden2011turbulence}, those of~\cite{mohammadnejad2020thick,zhou2017thin,zhou2015distributed,zhou2015simultaneous,dunn2007new,dunn2009compositional,dunn2010finite,wang2018direct,wang2017direct,zhou2015visualization,wang2019structure} suggest that the reaction zone is in fact broadened compared to the laminar flame counterpart. For example, in a recent investigation~\cite{mohammadnejad2020thick}, the authors show that for flames with $Ka \approx 76$, the reaction zone thickness can increase up to 3.9 times that of the corresponding laminar flame.  Mohammadnejad et al.~\cite{mohammadnejad2020thick} speculated that relatively small size energetic eddies may penetrate into and broaden the reaction zone. It was shown \cite{mohammadnejad2020thick} that the probability density function of eddy size distribution and kinetic energy of the eddies can be influenced by the type of the utilized turbulence generating mechanism, and this may be the reason for the discrepancy in the literature regarding the turbulent premixed flames reaction zone thickness.



The internal structure of turbulent premixed flames has implications for estimation of the turbulent burning velocity. This parameter can be estimated globally and locally, which are referred to as the global and local consumption speeds, respectively \cite{driscoll2008turbulent,driscoll2020premixed}. These speeds are estimated using \cite{wabel2017turbulent,driscoll2020premixed,driscoll2008turbulent}
\begin{subequations}
	\label{Eq:S_t}
	\begin{equation}
	\label{Eq:Glob}
	S_\mathrm{T,GC}=\frac{\dot{m}_\mathrm{r}}{\rho_\mathrm{r} \overline{A}_\mathrm{f}}
	\end{equation} 
	\begin{equation}
	\label{Eq:Loc}
	S_\mathrm{T,LC}=\frac{S_\mathrm{L,0} I_0}{L_\mathrm{\xi}} \int_{-\infty}^{\infty} \int_{-\infty}^{\infty}\Sigma(\eta,\xi) d\eta d\xi,
	\end{equation} 
\end{subequations}
where $S_\mathrm{T,GC}$ and $S_\mathrm{T,LC}$ are referred to as the global and local consumption speeds, respectively~\cite{driscoll2008turbulent}. $\dot{m}_\mathrm{r}$ and $\rho_\mathrm{r}$ are the reactants mass flow rate and density. $\overline{A}_\mathrm{f}$ is the flame surface area estimated based on the mean progress variable ($\overline{c}$). In Equation~(\ref{Eq:Loc}), $\Sigma$ is the flame surface density, $L_\mathrm{\xi}$ is the length of the mean progress variable contour, $I_0$ is the Bray-Cant \cite{bray1991some} stretch factor, and $\eta$ and $\xi$ are the curvilinear axes, which are respectively normal and tangent to a given $\overline{c}$ contour (usually $\overline{c} = 0.5$). As reviewed in Driscoll et al.~\cite{driscoll2008turbulent,driscoll2020premixed}, there are two ongoing questions regarding the global and local consumption speeds of methane-air turbulent premixed flames in the literature. First, the DNS studies of~\cite{lapointe2016fuel,nivarti2017scalar,nivarti2017direct} suggest that increasing the turbulence intensity increases the flame surface area such that the local consumption speed nearly equals the global consumption speed. However, several other studies, see for example ~\cite{driscoll2020premixed,nivarti2019reconciling,gulder2007contribution,wabel2017turbulent,lee2012validation,peters1999turbulent,wang2019structure,yuen2010dynamics}, suggest that as the turbulence intensity increases, the global consumption speed increases, but the flame surface area and the local consumption speed plateau. For instance, Wabel et al.~\cite{wabel2017turbulent} show that, at $u'/S_\mathrm{L,0} \approx 150$, the global consumption speed is about 5 times the local consumption speed. Wabel et al.~\cite{wabel2017turbulent} speculated that, as the preheat zone broadens with increasing the turbulence intensity, the turbulent diffusivity of the gas in this zone increases, enhancing the transport of reactants and as a result increase of the global consumption speed. Recently, Mohammadnejad et al.~\cite{mohammadnejad2020thick} showed that, in addition to the preheat zone, the reaction zone also broadens with increasing the turbulence intensity, which may also contribute to the increase of the gas diffusivity. Both G\"{u}lder~\cite{gulder2007contribution} and Nivarti et al.~\cite{nivarti2019reconciling} performed theoretical calculations and showed that the enhancement of gas diffusivity inside the flame region can allow for elaborating the reason for the increase of the global consumption speed with increase of the turbulence intensity and as a result the difference between the local and global consumption speed values reported in for example~\cite{driscoll2020premixed,nivarti2019reconciling,gulder2007contribution,wabel2017turbulent,lee2012validation,peters1999turbulent,wang2019structure,yuen2010dynamics}. However, in both G\"{u}lder~\cite{gulder2007contribution} and Nivarti et al.~\cite{nivarti2019reconciling}, the preheat and reaction zones thicknesses are not measured; and as a result, it is not known whether the enhancement of gas diffusivity is due to broadening of only the preheat zone or both the preheat and reaction zones. This requires further investigations.


The second question is related to the behavior of the global consumption speed itself. As hypothesized by Damk\"{o}hler~\cite{damkohler1940einfluss}, also known as the first Damk\"{o}hler's hypothesis, increasing $u'/S_\mathrm{L,0}$ is expected to increase the global consumption speed linearly. Although, at relatively small turbulence intensities, the global consumption speed nearly follows a linear relation with the turbulence intensity~\cite{driscoll2020premixed,kuo2012fundamentals,peters2000turbulent,damkohler1940einfluss,wabel2017turbulent,wang2019structure,yuen2010dynamics,yuen2013turbulent}, at relatively intense turbulence conditions, the global consumption speed plotted against the turbulence intensity bends towards the axis of $u'/S_\mathrm{L,0}$. This is referred to as the bending behavior in the literature, see for example~\cite{wabel2017turbulent,wang2019structure,yuen2010dynamics,yuen2013turbulent,kido2002influence,filatyev2005measured}. Several reasons such as nonlinear variation of gas diffusivity with turbulence intensity~\cite{driscoll2020premixed,damkohler1940einfluss,kuo2012fundamentals,nivarti2019reconciling} as well as local quenching \cite{driscoll2020premixed} are suggested to contribute to the bending behavior. 
To our best of knowledge, a potential relation between the flame quenching (or local extinction) and the bending behavior has not been studied yet.




\subsection{Effect of hydrogen enrichment on turbulent premixed flame internal structure and burning velocity}

The internal structure of hydrogen-enriched turbulent methane-air premixed flames has been recently investigated in Mohammadnejad et al.~\cite{mohammadnejad2019internal} and Zhang et al.~\cite{zhang2020effect}. The study of Mohammadnejad et al.~\cite{mohammadnejad2019internal} shows that, for relatively moderate turbulence conditions ($u'/S_\mathrm{L,0}$ up to 2.6), the preheat and reaction zone thicknesses of the hydrogen-enriched flames are similar to the laminar premixed flame counterparts. Mohammadnejad et al.~\cite{mohammadnejad2019internal} showed that hydrogen enrichment (up to 50\%) does not significantly influence the preheat and reaction zone thicknesses normalized by the corresponding laminar flame counterparts. Compared to~\cite{mohammadnejad2019internal}, whose tested turbulence intensity was limited to 2.6, Zhang et al.~\cite{zhang2020effect} investigated both the reaction and preheat zone thicknesses for larger turbulence intensities ($3 \lesssim u'/S_\mathrm{L,0} \lesssim 12.5$) with hydrogen-enrichment percentages examined up to 60\%. They~\cite{zhang2020effect} showed that, similar to the methane-air turbulent premixed flames, increasing the turbulence intensity can lead to broadening of the hydrogen-enriched methane-air flames preheat zone. However, Zhang et al.~\cite{zhang2020effect} showed that the reaction zone thickness does not change by increasing the turbulence intensity. Although studies of both~\cite{mohammadnejad2019internal,zhang2020effect} provide insight into the structure of hydrogen-enriched methane-air premixed flames at relatively small and moderate turbulence intensities, the preheat and reaction zone thicknesses of these flames pertaining to relatively intense turbulence conditions remain to be investigated.

Few experimental investigations~\cite{guo2010burning,halter2007characterization} have studied the effect of hydrogen-enrichment on local and global consumption speeds of turbulent premixed flames. Both Halter et al.~\cite{halter2007characterization} and Guo et al.~\cite{guo2010burning} showed that increasing the hydrogen-enrichment percentage increases the global consumption speed normalized by the unstretched laminar flame speed. Guo et al.~\cite{guo2010burning} showed this parameter also increases with increasing the turbulence intensity. The values of the local consumption speed are not reported in~\cite{halter2007characterization}, however, Guo et al.~\cite{guo2010burning} showed that increasing both the hydrogen-enrichment percentage and the turbulence intensity increase the local consumption speed normalized by the unstretched laminar flame speed. Although these studies~\cite{guo2010burning,halter2007characterization} elaborated the effect of hydrogen enrichment and turbulence intensity on the local/global consumption speeds, their investigations are performed for $u'/S_\mathrm{L,0} \lesssim 5$. To our best of knowledge, values of the local and global consumption speeds for hydrogen-enriched premixed flame are not investigated at larger turbulence intensities.


This study aims to address three objectives. The first objective of the present study is to measure the preheat and reaction zone thicknesses of hydrogen-enriched methane-air premixed flames pertaining to relatively intense turbulence conditions. To our best of knowledge, such information is not available in the literature. The second objective of this study is to develop a consumption speed estimation framework that does not rely on the flamelet assumption. Then, using this framework, we aim to understand the reason for the difference between local and global consumption speeds reported in the literature. The third objective of the present study is to characterize the bending behavior of the global consumption speed and to develop a mathematical formulation that allows to estimate the amount of bending.


\section{Experimental methodology}
\label{EM}

The experimental setup, diagnostics, coordinate system, and the tested conditions are presented in this section.

\subsection{Setup} 

\begin{figure*}
	\centering
	\includegraphics[width = 1.0\textwidth]{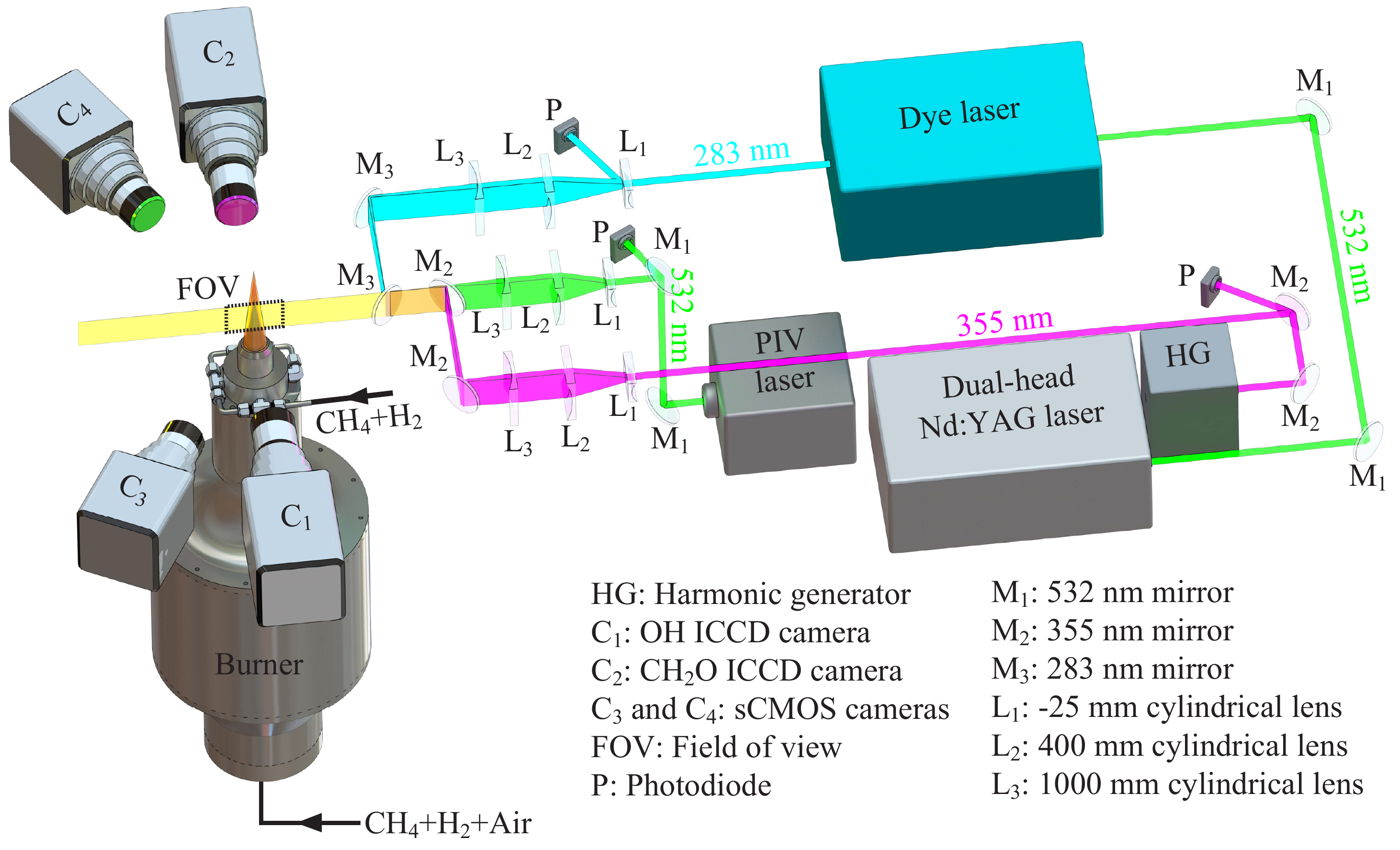} 
	\caption{Schematic of the experimental setup and the diagnostics.}
	\label{fig:Setup}
\end{figure*}

\begin{figure*}
	\centering
	\includegraphics[width = 0.7\textwidth]{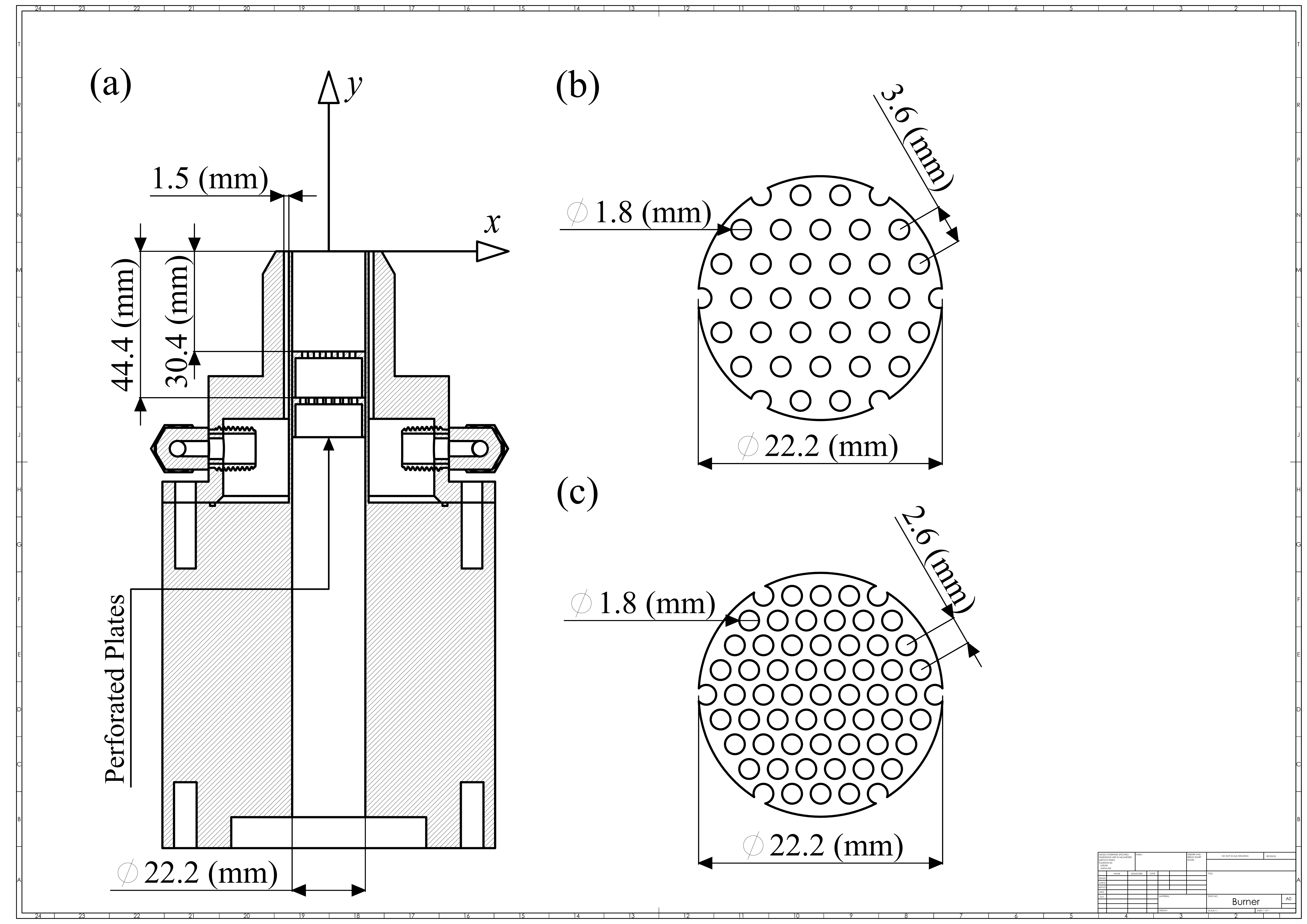} 
	\caption{Technical drawing of (a) the burner nozzle as well as (b and c) the perforated plates.}
	\label{fig:Burner}
\end{figure*}

The setup used in this study is a Bunsen burner identical to that utilized in~\cite{mohammadnejad2020thick,tamadonfar2015experimental}. This burner consists of an expansion section, a settling chamber (with 5 mesh screens), a contraction section, and a nozzle with exit diameter of 22.2~mm. The burner schematic and the burner nozzle technical drawing are shown in Figs.~\ref{fig:Setup}~and~\ref{fig:Burner}(a), respectively. A mixture of 75\% hydrogen and 25\% methane (volumetric percentages) is injected through the burner rim, generating a pilot flame, see Fig.~\ref{fig:Setup}. Similar to \cite{mohammadnejad2020thick}, the volumetric flowrate of the pilot flow is only 2\% of the main flow, which is small, and as a result, the pilot flame is not expected to influence the main Bunsen flames. In order to generate a relatively wide range of turbulence intensities, three different turbulence generating mechanisms are used similar to~\cite{mohammadnejad2020thick}. Specifically, zero, one, or two perforated plates were utilized. For the conditions with no perforated plate (which is the first turbulence generating mechanism), turbulence is primarily produced by the mesh screens and the Kelvin-Helmholtz instability that is developed in the jet shear layer, similar to~\cite{mohammadnejad2020thick,kheirkhah2014topology,kheirkhah2016periodic,kheirkhah2016experimental}. For the second turbulence generating mechanism, a perforated plate, with the technical drawing shown in Fig.~\ref{fig:Burner}(c), is placed 44.4~mm upstream of the nozzle exit plane. This plate contains 1.8~mm diameter circular holes arranged in a hexagonal pattern, generating a blockage ratio of $60$\%. For the third turbulence generating mechanism, one perforated plate (identical to that used for the second turbulence generating mechanism) is positioned 44.4~mm, and a second plate is located 30.4~mm upstream of the burner exit plane, see Fig.~\ref{fig:Burner}(b). The second perforated plate has holes positioned on a hexagonal arrangement with a diameter of 1.8~mm. The blockage ratio of this plate is $77$\%.

\subsection{Diagnostics} 

Simultaneous Planar Laser-Induced Fluorescence (PLIF) of hydroxyl radical ($\mathrm{OH}$) and formaldehyde molecule ($\mathrm{CH_2O}$) as well as separate Stereoscopic Particle Image Velocimetry (SPIV) were performed in this study. The diagnostics used here is similar to that used in Mohammadnejad et al.~\cite{mohammadnejad2020thick}. However, compared to \cite{mohammadnejad2020thick}, the SPIV experiments in the present study are performed only for the non-reacting flow conditions and to characterize the background turbulent flow. Both PLIF and SPIV data acquisition rate were 1~Hz. A dual-head Nd:YAG pump laser (Quanta Ray PIV400) and a frequency-doubled dye laser (Sirah Precision Scan) are used for generating the PLIF signals. Two laser pulses with wavelengths of 532~nm and $355\pm3$~nm are produced by the pump laser using harmonic generators. As shown in Fig.~\ref{fig:Setup}, the 532~nm beam is used to pump the dye laser to generate a 283 nm ($282.94\pm0.005$~nm) beam. This beam is used to excite $\mathrm{Q_1}$(6) line of the OH A-X system (1,0) vibrational band~\cite{mohammadnejad2020thick,mohammadnejad2019internal,fayoux2005experimental,ayoola2006spatially,roder2012simultaneous,roder2013simultaneous,hardalupas2010spatial,paul1998planar,ayoola2009temperature,vena2015heat}. The 355~nm beam is used to excite the $\tilde{A}^1A_2-\tilde{X}^1A_14_0^1$ vibronic manifold of $\mathrm{CH_2O}$ similar to~\cite{mohammadnejad2020thick,mohammadnejad2019internal,fayoux2005experimental,roder2012simultaneous,roder2013simultaneous,hardalupas2010spatial,harrington1993laser,brackmann2003laser,yamamoto2011local,vena2015heat}. Using mirrors ($\mathrm{M_1-M_3}$) and cylindrical lenses ($\mathrm{L_1-L_3}$) shown in Fig.~\ref{fig:Setup}, the 282.94~nm and 355~nm beams are converted to coincident laser sheets with the thickness and height of about $250~\mathrm{\mu m}$ and 40~mm, respectively. Two PIMAX ICCD cameras (see, $\mathrm{C_1}$ and $\mathrm{C_2}$ in Fig.~\ref{fig:Setup}), each equipped with a Nikkor UV lens and a bandpass filter with center wavelength of $320\pm20$~nm (for OH) and Coastal Optics UV lens and a Schott GG~395 longpass filter (for $\mathrm{CH_2O}$), are used to collect the PLIF signals. The projected spatial resolution of both cameras is $89~\mathrm{\mu m}$ per pixel. The knife edge technique \cite{clemens2002flow,wang2004effects} was used to obtain the line spread functions of both cameras. It was calculated that the optical system effective resolution (the full width at half maximum of the line-spread function) is $263~\mathrm{\mu m}$. Two photodiodes, shown in Fig.~\ref{fig:Setup}, are utilized to collect pulse-to-pulse variations of the 282.94~nm and 355~nm beams energies, which are utilized for reduction of the PLIF data discussed in the next section. Further details regarding the utilized PLIF system can be found in~\cite{mohammadnejad2020thick,mohammadnejad2019internal}.

The SPIV system contains a dual-cavity, double-pulse, Nd:YAG laser (BSL Twins CFR PIV200) as well as $\mathrm{C_3}$ and $\mathrm{C_4}$ sCMOS cameras (LaVision Imager sCMOS), which are equipped with Scheimpflug adapters and Tokina lenses, see Fig.~\ref{fig:Setup}. The laser generates pairs of 532~nm beams with a 4--33~$\mu$s separation time (depending on the tested mean bulk flow velocity). $\mathrm{M_1-M_3}$ mirrors as well as $\mathrm{L_1-L_3}$ cylindrical lenses (see Fig.~\ref{fig:Setup}) are utilized to convert the PIV laser beams to laser sheets that are coincident with those of the PLIF lasers. The field of view and projected spatial resolution of the SPIV cameras are 84~mm$\times$70~mm and 29.3~$\mu$m/pixel, respectively. Olive oil is atomized using a TSI Six-Jet atomizer and is fed into the burner. PLIF and SPIV images are registered and mapped to one physical coordinate system using a three-dimensional LaVision Type~20 target plate. LaVision DaVis~8.4 software is used to calculate the velocity vectors. The details of the PIV vector calculation is identical to that described in~\cite{mohammadnejad2020thick}.

\subsection{Coordinate system} 

\begin{figure*}
	\centering
	\includegraphics[width = 0.4\textwidth]{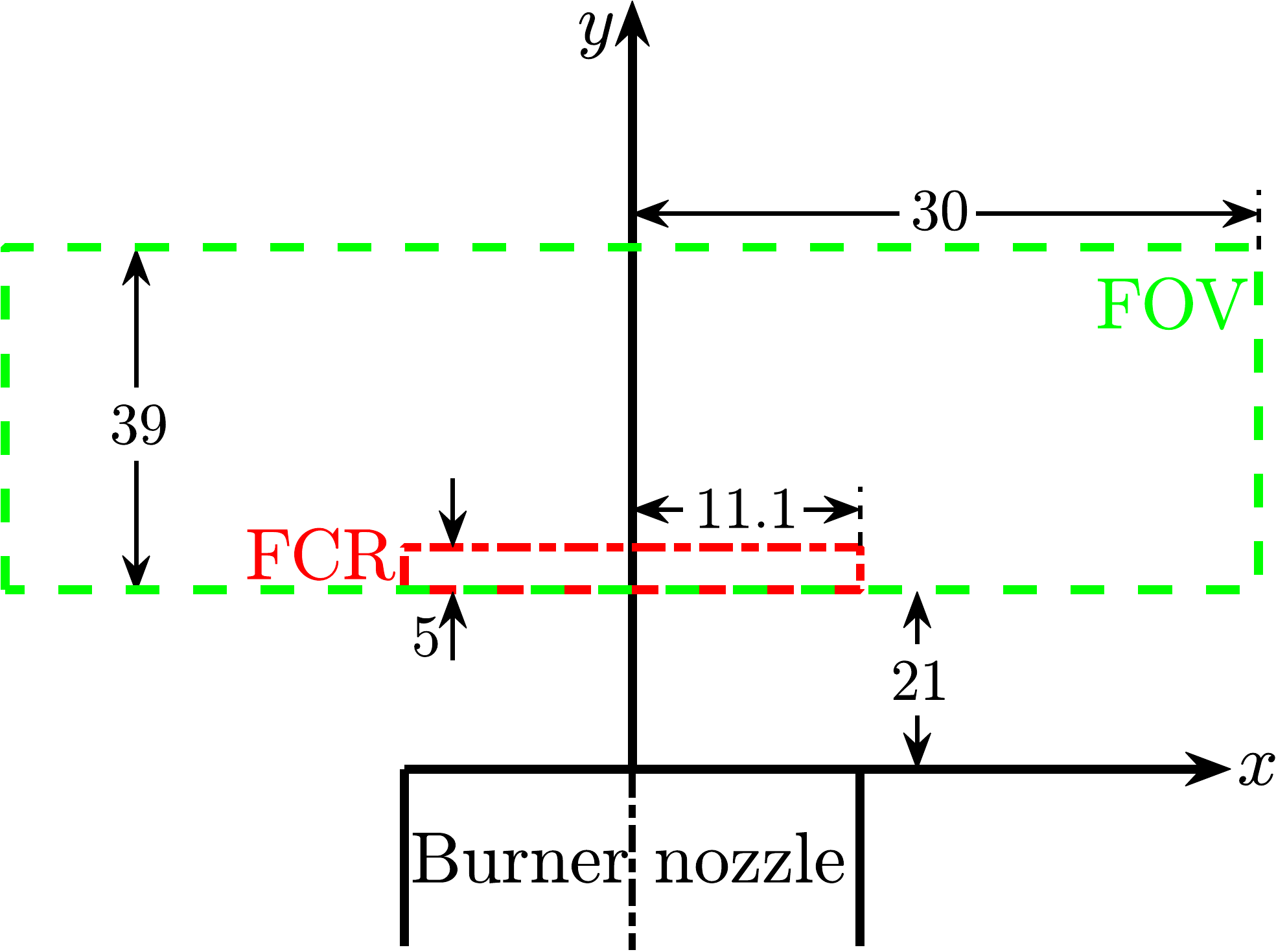}
	\caption{The coordinate system, field of view, and flow characterization region (FCR). The dimensions are in mm.}
	\label{fig:Coordinates}
\end{figure*}

A Cartesian coordinate system, presented in Fig.~\ref{fig:Coordinates}, is used in the present study. The origin of the coordinate system is at the intersection of the burner centerline and the nozzle exit plane. The $x-$ and $y-$axes are at the imaging plane and lie inside the laser sheets. The $z-$axis is normal to the $x-y$ plane. The PLIF and SPIV field of views, shown by a green dashed window in Fig.~\ref{fig:Coordinates}, are coincident. In the figure, the flow characterization region (FCR) highlights the location at which the non-reacting turbulent flow characteristics are averaged, which is similar to~\cite{mohammadnejad2020thick}.

\subsection{Tested conditions} 
In total, 68 experimental conditions are tested, with the corresponding details tabulated in Table~\ref{tab:Conditions}. Mixture of methane (grade~2 with 99\% chemical purity) and hydrogen (grade~5 with 99.999\% chemical purity) is used as the fuel in this study. The fuel-air equivalence ratio ($\phi$) for all tested conditions is 0.7. All tested conditions labels are provided in the first column of the table. In this column, U$\#$ and H$\#$ indicate the tested mean bulk velocity ($U$) and the hydrogen-enrichment percentage ($\mathrm{H_2}$\%). Four mean bulk flow velocities of 5, 15, 25, and 35~m/s are examined, which are provided in the second column of the table. $\mathrm{H_2}\%$ varied from 0 to a maximum of $70$\%, with the corresponding values provided in the third column of Table~\ref{tab:Conditions}. In the label of each tested condition, T$\#$ indicates the number of utilized perforated plates. Conditions with zero (T0), one (T1), and two (T2) perforated plates are highlighted by the blue, green, and red colors in the table, respectively. Cantera simulations \cite{cantera} with the GRI-Mech~3.0 mechanism are used to estimate the unstretched laminar flame speeds ($S_\mathrm{L,0}$). Laminar flame thickness is calculated using $\delta_\mathrm{L}=(\lambda/c_\mathrm{p})/(\rho_0S_\mathrm{L,0})$, with $\lambda$ and $c_\mathrm{p}$ being the gas thermal conductivity and specific heat (both estimated at 1500~K), and $\rho_0$ being the fuel and air mixture density (estimated at 300~K), similar to \cite{mohammadnejad2020thick,driscoll2020premixed,skiba2018premixed,peters2000turbulent}. 

Root Mean Square (RMS) of the velocity fluctuations along the $y$ ($u'$), $x$ ($v'$), and $z$ ($w'$) directions as well as the integral ($\Lambda$), Taylor ($l_\mathrm{T}=\Lambda (u^\prime \Lambda/ \nu)^{-1/2}$), and Kolmogorov ($\eta_\mathrm{K}=(\nu^3/\epsilon)^{1/4}=\Lambda (u^\prime \Lambda/ \nu)^{-3/4}$) length scales are averaged within the red dotted-dashed window of Fig.~\ref{fig:Coordinates}, and the results are presented in the table. In the equations used for calculation of the Taylor and Kolmogorov length scales, $\nu$ is the fuel and air mixture kinematic viscosity (estimated at 300~K) and $\epsilon=u'^3/\Lambda$~\cite{driscoll2020premixed,pope2001turbulent}. The integral length scale, $\Lambda$, is estimated using the formulations provided in~\cite{pope2001turbulent} and the procedures discussed in~\cite{mohammadnejad2020thick,mohammadnejad2019internal}. Similar to~\cite{mohammadnejad2020thick}, conditions pertaining to the first turbulence generating mechanism feature the Kelvin-Helmholtz instability. This instability is generated by the jet shear layer, can influence the values of the RMS velocity and integral length scale, and does not correspond to the background turbulent flow. Thus, using the Proper Orthogonal Decomposition technique~\cite{meyer2007turbulent} and similar to~\cite{mohammadnejad2020thick}, the Kelvin-Helmholtz modes are identified and excluded from the non-reacting flow data pertaining to zero perforated plates, following the procedures discussed in~\cite{mohammadnejad2020thick}. The Reynolds ($Re_\mathrm{T}$) and Karlovitz ($Ka$) numbers are estimated using $Re_\mathrm{T} =u^\prime \Lambda/ \nu$ and $Ka = (u'/S_\mathrm{L,0})^{3/2}(\Lambda/\delta_\mathrm{L})^{-1/2}$, with the values provided in Table~\ref{tab:Conditions}. The effective Lewis number ($L_\mathrm{eff}$) of the fuel and air mixture was estimated using the formulations provided in Mohammadnejad et al.~\cite{mohammadnejad2019internal}, with the corresponding values presented in the last column of the table.

\begin{figure}
	\centering
	\includegraphics[width = 0.9\textwidth]{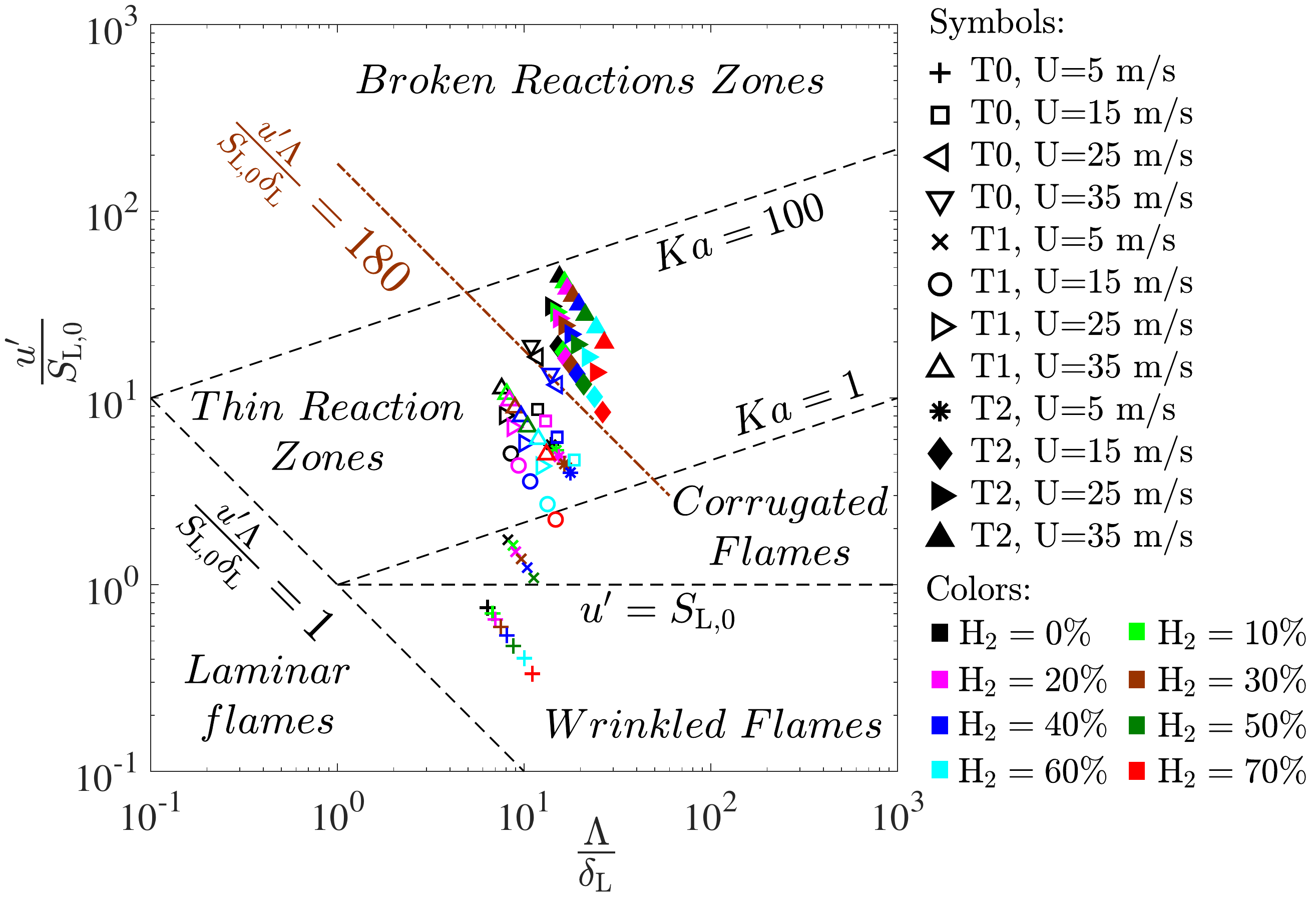} 
	\caption{The tested experimental conditions overlaid on the Borghi--Peters diagram \cite{peters2000turbulent}.}
	\label{fig:Borghi}
\end{figure}

All test conditions are overlaid on the Borghi--Peters~\cite{peters2000turbulent,borghi1988turbulent} diagram, as shown in Fig.~\ref{fig:Borghi}. The results in the figure suggest that flames with $U=5$~m/s and no perforated plate pertain to the wrinkled flames regime. Flames with $U=5$~m/s and one perforated plate as well as Flame~U15H70T1 pertain to the corrugated flames regime. The rest of the tested conditions are positioned in the thin reaction zones regime. In Fig.~\ref{fig:Borghi}, the newly-suggested border of $u'\Lambda/(S_\mathrm{L,0}\delta_\mathrm{L}) = 180$~\cite{skiba2018premixed,driscoll2020premixed}, which separates the flames with thin and broadened preheat zones, is also presented. This border suggests that the conditions pertaining to $U=15, 25$, and 35~m/s with 2 perforated plates as well as Flame~U35H00T0 are expected to feature a broadened preheat zone. Flames~U25H00T0, U25H40T0, and U35H40T0 are positioned near the border of $u'\Lambda/(S_\mathrm{L,0}\delta_\mathrm{L}) = 180$. The rest of the tested conditions are expected to feature a relatively thin preheat zone. The predictions of the Borghi-Peters diagram related to thin and broadened preheat zones are discussed in detail later in the results section.

\begin{table}[!htbp]
	\caption{Tested experimental conditions. For all conditions, $\phi=0.7$. $U$, $S_\mathrm{L,0}$, $u'$, $v'$, and $w'$ are in m/s and $\delta_\mathrm{L}$, $\eta_\mathrm{K}$, $l_\mathrm{T}$ and $\Lambda$ are in mm.}
	\label{tab:Conditions}
	\centering
	\scalebox{0.71}{
		\begin{tabular}{ccccccccccccccccc}
			\hline
			& $U$ & $\mathrm{H_2}\%$ & $S_\mathrm{L,0}$ & $\delta_\mathrm{L}$ & $u'$ & $v'$ & $w'$ & $\eta_\mathrm{K}$& $l_\mathrm{T}$ & $\Lambda$ & $u'/U$ & $u'/S_\mathrm{L,0}$ & $\Lambda/\delta_\mathrm{L}$ & $Re_\mathrm{T}$ & $Ka$ & $Le_\mathrm{eff}$ \\
			\hline
			\color{blue}U5H00T0 & 5 & 0 & 0.19 & 0.33 & 0.15 & 0.04 & 0.07 & 0.23 & 0.48 & 2.1 & 0.03 & 0.8 & 6.4 & 19 & 0.3 & 1.03 \\ 
			\hline
			\color{blue}U5H10T0 & 5 & 10 & 0.21 & 0.31 & 0.15 & 0.04 & 0.07 & 0.23 & 0.48 & 2.1 & 0.03 & 0.7 & 6.7 & 19 & 0.2 & 1.00 \\ 
			\hline
			\color{blue}U5H20T0 & 5 & 20 & 0.23 & 0.30 & 0.15 & 0.04 & 0.07 & 0.23 & 0.48 & 2.1 & 0.03 & 0.7 & 7.1 & 19 & 0.2 & 0.97 \\ 
			\hline
			\color{blue}U5H30T0 & 5 & 30 & 0.25 & 0.28 & 0.15 & 0.04 & 0.07 & 0.23 & 0.49 & 2.1 & 0.03 & 0.6 & 7.5 & 19 & 0.2 & 0.93 \\ 
			\hline
			\color{blue}U5H40T0 & 5 & 40 & 0.27 & 0.26 & 0.15 & 0.04 & 0.07 & 0.24 & 0.49 & 2.1 & 0.03 & 0.5 & 8.1 & 19 & 0.1 & 0.91 \\ 
			\hline
			\color{blue}U5H50T0 & 5 & 50 & 0.31 & 0.24 & 0.15 & 0.04 & 0.07 & 0.24 & 0.49 & 2.1 & 0.03 & 0.5 & 8.8 & 18 & 0.1 & 0.88 \\ 
			\hline
			\color{blue}U5H60T0 & 5 & 60 & 0.36 & 0.21 & 0.15 & 0.04 & 0.07 & 0.24 & 0.49 & 2.1 & 0.03 & 0.4 & 9.8 & 18 & 0.1 & 0.87 \\
			\hline
			\color{blue}U5H70T0 & 5 & 70 & 0.44 & 0.19 & 0.15 & 0.04 & 0.07 & 0.24 & 0.50 & 2.1 & 0.03 & 0.3 & 11.2 & 18 & 0.1 & 0.86 \\ 
			\hline
			\color{blue}U15H00T0 & 15 & 0 & 0.19 & 0.33 & 1.70 & 1.36 & 0.54 & 0.04 & 0.19 & 3.9 & 0.11 & 8.7 & 11.9 & 410 & 7.4 & 1.03 \\ 
			\hline
			\color{blue}U15H20T0 & 15 & 20 & 0.23 & 0.30 & 1.70 & 1.36 & 0.54 & 0.04 & 0.19 & 3.9 & 0.11 & 7.5 & 13.1 & 405 & 5.7 & 0.97 \\ 
			\hline
			\color{blue}U15H40T0 & 15 & 40 & 0.27 & 0.26 & 1.70 & 1.36 & 0.54 & 0.04 & 0.20 & 3.9 & 0.11 & 6.2 & 15.0 & 397 & 3.9 & 0.91 \\ 
			\hline
			\color{blue}U15H60T0 & 15 & 60 & 0.36 & 0.21 & 1.70 & 1.36 & 0.54 & 0.04 & 0.20 & 3.9 & 0.11 & 4.7 & 18.2 & 387 & 2.4 & 0.87 \\ 
			\hline
			\color{blue}U25H00T0 & 25 & 0 & 0.19 & 0.33 & 3.24 & 2.46 & 1.26 & 0.03 & 0.14 & 3.9 & 0.13 & 16.6 & 11.9 & 783 & 19.6 & 1.03 \\ 
			\hline
			\color{blue}U25H40T0 & 25 & 40 & 0.27 & 0.26 & 3.24 & 2.46 & 1.26 & 0.03 & 0.14 & 3.9 & 0.13 & 11.8 & 15.0 & 759 & 10.4 & 0.91 \\ 
			\hline
			\color{blue}U35H00T0 & 35 & 0 & 0.19 & 0.33 & 3.72 & 2.96 & 1.83 & 0.02 & 0.13 & 3.6 & 0.11 & 19.1 & 11.0 & 829 & 25.1 & 1.03 \\ 
			\hline
			\color{blue}U35H40T0 & 35 & 40 & 0.27 & 0.26 & 3.72 & 2.96 & 1.83 & 0.02 & 0.13 & 3.6 & 0.11 & 13.5 & 13.9 & 804 & 13.3 & 0.91 \\ 
			\hline
			\color{green}U5H00T1 & 5 & 0 & 0.19 & 0.33 & 0.34 & 0.25 & 0.30 & 0.13 & 0.36 & 2.7 & 0.07 & 1.7 & 8.3 & 57 & 0.8 & 1.03 \\ 
			\hline
			\color{green}U5H10T1 & 5 & 10 & 0.21 & 0.31 & 0.34 & 0.25 & 0.30 & 0.13 & 0.36 & 2.7 & 0.07 & 1.6 & 8.7 & 57 & 0.7 & 1.00 \\ 
			\hline
			\color{green}U5H20T1 & 5 & 20 & 0.23 & 0.30 & 0.34 & 0.25 & 0.30 & 0.13 & 0.36 & 2.7 & 0.07 & 1.5 & 9.1 & 56 & 0.6 & 0.97 \\ 
			\hline
			\color{green}U5H30T1 & 5 & 30 & 0.25 & 0.28 & 0.34 & 0.25 & 0.30 & 0.13 & 0.36 & 2.7 & 0.07 & 1.4 & 9.7 & 56 & 0.5 & 0.93 \\ 
			\hline
			\color{green}U5H40T1 & 5 & 40 & 0.27 & 0.26 & 0.34 & 0.25 & 0.30 & 0.13 & 0.36 & 2.7 & 0.07 & 1.2 & 10.5 & 55 & 0.4 & 0.91 \\ 
			\hline
			\color{green}U5H50T1 & 5 & 50 & 0.31 & 0.24 & 0.34 & 0.25 & 0.30 & 0.13 & 0.37 & 2.7 & 0.07 & 1.1 & 11.4 & 55 & 0.3 & 0.88 \\ 
			\hline
			\color{green}U15H00T1 & 15 & 0 & 0.19 & 0.33 & 0.98 & 0.71 & 0.83 & 0.06 & 0.22 & 2.8 & 0.07 & 5.0 & 8.6 & 173 & 3.9 & 1.03 \\ 
			\hline
			\color{green}U15H20T1 & 15 & 20 & 0.23 & 0.30 & 0.98 & 0.71 & 0.83 & 0.06 & 0.22 & 2.8 & 0.07 & 4.4 & 9.5 & 171 & 2.9 & 0.97 \\ 
			\hline
			\color{green}U15H40T1 & 15 & 40 & 0.27 & 0.26 & 0.98 & 0.71 & 0.83 & 0.06 & 0.22 & 2.8 & 0.07 & 3.6 & 10.9 & 168 & 2.0 & 0.91 \\ 
			\hline
			\color{green}U15H60T1 & 15 & 60 & 0.36 & 0.21 & 0.98 & 0.71 & 0.83 & 0.06 & 0.22 & 2.8 & 0.07 & 2.7 & 13.2 & 163 & 1.2 & 0.87 \\ 
			\hline
			\color{green}U15H70T1 & 15 & 70 & 0.44 & 0.19 & 0.98 & 0.71 & 0.83 & 0.06 & 0.22 & 2.8 & 0.07 & 2.2 & 15.1 & 160 & 0.9 & 0.86 \\ 
			\hline
			\color{green}U25H00T1 & 25 & 0 & 0.19 & 0.33 & 1.57 & 1.17 & 1.36 & 0.04 & 0.16 & 2.6 & 0.06 & 8.1 & 7.9 & 253 & 8.2 & 1.03 \\ 
			\hline
			\color{green}U25H20T1 & 25 & 20 & 0.23 & 0.30 & 1.57 & 1.17 & 1.36 & 0.04 & 0.16 & 2.6 & 0.06 & 7.0 & 8.7 & 250 & 6.2 & 0.97 \\ 
			\hline
			\color{green}U25H40T1 & 25 & 40 & 0.27 & 0.26 & 1.57 & 1.17 & 1.36 & 0.04 & 0.17 & 2.6 & 0.06 & 5.7 & 10.0 & 245 & 4.3 & 0.91 \\ 
			\hline
			\color{green}U25H60T1 & 25 & 60 & 0.36 & 0.21 & 1.57 & 1.17 & 1.36 & 0.04 & 0.17 & 2.6 & 0.06 & 4.3 & 12.1 & 239 & 2.6 & 0.87 \\ 
			\hline
			\color{green}U35H00T1 & 35 & 0 & 0.19 & 0.33 & 2.19 & 1.71 & 1.94 & 0.03 & 0.14 & 2.5 & 0.06 & 11.3 & 7.6 & 341 & 13.7 & 1.03 \\ 
			\hline
			\color{green}U35H10T1 & 35 & 10 & 0.21 & 0.31 & 2.19 & 1.71 & 1.94 & 0.03 & 0.14 & 2.5 & 0.06 & 10.5 & 8.0 & 339 & 12.0 & 1.00 \\ 
			\hline
			\color{green}U35H20T1 & 35 & 20 & 0.23 & 0.30 & 2.19 & 1.71 & 1.94 & 0.03 & 0.14 & 2.5 & 0.06 & 9.7 & 8.4 & 337 & 10.4 & 0.97 \\ 
			\hline
			\color{green}U35H30T1 & 35 & 30 & 0.25 & 0.28 & 2.19 & 1.71 & 1.94 & 0.03 & 0.14 & 2.5 & 0.06 & 8.9 & 9.0 & 334 & 8.8 & 0.93 \\ 
			\hline
			\color{green}U35H40T1 & 35 & 40 & 0.27 & 0.26 & 2.19 & 1.71 & 1.94 & 0.03 & 0.14 & 2.5 & 0.06 & 8.0 & 9.7 & 331 & 7.3 & 0.91 \\ 
			\hline
			\color{green}U35H50T1 & 35 & 50 & 0.31 & 0.24 & 2.19 & 1.71 & 1.94 & 0.03 & 0.14 & 2.5 & 0.06 & 7.0 & 10.6 & 327 & 5.7 & 0.88 \\ 
			\hline
			\color{green}U35H60T1 & 35 & 60 & 0.36 & 0.21 & 2.19 & 1.71 & 1.94 & 0.03 & 0.14 & 2.5 & 0.06 & 6.0 & 11.7 & 322 & 4.3 & 0.87 \\ 
			\hline
			\color{green}U35H70T1 & 35 & 70 & 0.44 & 0.19 & 2.19 & 1.71 & 1.94 & 0.03 & 0.14 & 2.5 & 0.06 & 5.0 & 13.3 & 316 & 3.1 & 0.86 \\ 
			\hline
			\color{red}U5H00T2 & 5 & 0 & 0.19 & 0.33 & 1.10 & 0.84 & 0.89 & 0.06 & 0.26 & 4.6 & 0.22 & 5.6 & 13.9 & 311 & 3.6 & 1.03 \\ 
			\hline
			\color{red}U5H10T2 & 5 & 10 & 0.21 & 0.31 & 1.10 & 0.84 & 0.89 & 0.06 & 0.26 & 4.6 & 0.22 & 5.3 & 14.6 & 309 & 3.1 & 1.00 \\ 
			\hline
			\color{red}U5H20T2 & 5 & 20 & 0.23 & 0.30 & 1.10 & 0.84 & 0.89 & 0.06 & 0.26 & 4.6 & 0.22 & 4.9 & 15.4 & 307 & 2.7 & 0.97 \\ 
			\hline
			\color{red}U5H30T2 & 5 & 30 & 0.25 & 0.28 & 1.10 & 0.84 & 0.89 & 0.06 & 0.26 & 4.6 & 0.22 & 4.4 & 16.4 & 304 & 2.3 & 0.93 \\ 
			\hline
			\color{red}U5H40T2 & 5 & 40 & 0.27 & 0.26 & 1.10 & 0.84 & 0.89 & 0.06 & 0.26 & 4.6 & 0.22 & 4.0 & 17.6 & 301 & 1.9 & 0.91 \\ 
			\hline
			\color{red}U15H00T2 & 15 & 0 & 0.19 & 0.33 & 3.70 & 2.80 & 2.76 & 0.03 & 0.15 & 5.0 & 0.25 & 19.0 & 15.3 & 1151 & 21.1 & 1.03 \\ 
			\hline
			\color{red}U15H10T2 & 15 & 10 & 0.21 & 0.31 & 3.70 & 2.80 & 2.76 & 0.03 & 0.15 & 5.0 & 0.25 & 17.7 & 16.0 & 1144 & 18.6 & 1.00 \\ 
			\hline
			\color{red}U15H20T2 & 15 & 20 & 0.23 & 0.30 & 3.70 & 2.80 & 2.76 & 0.03 & 0.15 & 5.0 & 0.25 & 16.4 & 16.9 & 1136 & 16.1 & 0.97 \\ 
			\hline
			\color{red}U15H30T2 & 15 & 30 & 0.25 & 0.28 & 3.70 & 2.80 & 2.76 & 0.03 & 0.15 & 5.0 & 0.25 & 15.0 & 18.0 & 1127 & 13.6 & 0.93 \\ 
			\hline
			\color{red}U15H40T2 & 15 & 40 & 0.27 & 0.26 & 3.70 & 2.80 & 2.76 & 0.03 & 0.15 & 5.0 & 0.25 & 13.4 & 19.4 & 1116 & 11.2 & 0.91 \\ 
			\hline
			\color{red}U15H50T2 & 15 & 50 & 0.31 & 0.24 & 3.70 & 2.80 & 2.76 & 0.03 & 0.15 & 5.0 & 0.25 & 11.9 & 21.1 & 1103 & 8.9 & 0.88 \\ 
			\hline
			\color{red}U15H60T2 & 15 & 60 & 0.36 & 0.21 & 3.70 & 2.80 & 2.76 & 0.03 & 0.15 & 5.0 & 0.25 & 10.2 & 23.4 & 1086 & 6.7 & 0.87 \\ 
			\hline
			\color{red}U15H70T2 & 15 & 70 & 0.44 & 0.19 & 3.70 & 2.80 & 2.76 & 0.03 & 0.15 & 5.0 & 0.25 & 8.4 & 26.7 & 1065 & 4.7 & 0.86 \\ 
			\hline
			\color{red}U25H00T2 & 25 & 0 & 0.19 & 0.33 & 6.04 & 4.75 & 4.72 & 0.02 & 0.11 & 4.6 & 0.24 & 31.0 & 14.1 & 1730 & 45.9 & 1.03 \\ 
			\hline
			\color{red}U25H10T2 & 25 & 10 & 0.21 & 0.31 & 6.04 & 4.75 & 4.72 & 0.02 & 0.11 & 4.6 & 0.24 & 28.9 & 14.8 & 1720 & 40.5 & 1.00 \\ 
			\hline
			\color{red}U25H20T2 & 25 & 20 & 0.23 & 0.30 & 6.04 & 4.75 & 4.72 & 0.02 & 0.11 & 4.6 & 0.24 & 26.8 & 15.6 & 1708 & 35.1 & 0.97 \\ 
			\hline
			\color{red}U25H30T2 & 25 & 30 & 0.25 & 0.28 & 6.04 & 4.75 & 4.72 & 0.02 & 0.11 & 4.6 & 0.24 & 24.4 & 16.6 & 1694 & 29.7 & 0.93 \\ 
			\hline
			\color{red}U25H40T2 & 25 & 40 & 0.27 & 0.26 & 6.04 & 4.75 & 4.72 & 0.02 & 0.11 & 4.6 & 0.24 & 22.0 & 17.8 & 1677 & 24.4 & 0.91 \\ 
			\hline
			\color{red}U25H50T2 & 25 & 50 & 0.31 & 0.24 & 6.04 & 4.75 & 4.72 & 0.02 & 0.11 & 4.6 & 0.24 & 19.4 & 19.4 & 1657 & 19.3 & 0.88 \\ 
			\hline
			\color{red}U25H60T2 & 25 & 60 & 0.36 & 0.21 & 6.04 & 4.75 & 4.72 & 0.02 & 0.11 & 4.6 & 0.24 & 16.6 & 21.6 & 1632 & 14.6 & 0.87 \\ 
			\hline
			\color{red}U25H70T2 & 25 & 70 & 0.44 & 0.19 & 6.04 & 4.75 & 4.72 & 0.02 & 0.12 & 4.6 & 0.24 & 13.8 & 24.6 & 1600 & 10.3 & 0.86 \\ 
			\hline
			\color{red}U35H00T2 & 35 & 0 & 0.19 & 0.33 & 8.70 & 6.79 & 6.80 & 0.01 & 0.10 & 5.1 & 0.25 & 44.6 & 15.4 & 2729 & 76.0 & 1.03 \\ 
			\hline
			\color{red}U35H10T2 & 35 & 10 & 0.21 & 0.31 & 8.70 & 6.79 & 6.80 & 0.01 & 0.10 & 5.1 & 0.25 & 41.7 & 16.1 & 2713 & 67.0 & 1.00 \\ 
			\hline
			\color{red}U35H20T2 & 35 & 20 & 0.23 & 0.30 & 8.70 & 6.79 & 6.80 & 0.01 & 0.10 & 5.1 & 0.25 & 38.6 & 17.0 & 2695 & 58.0 & 0.97 \\ 
			\hline
			\color{red}U35H30T2 & 35 & 30 & 0.25 & 0.28 & 8.70 & 6.79 & 6.80 & 0.01 & 0.10 & 5.1 & 0.25 & 35.2 & 18.1 & 2673 & 49.1 & 0.93 \\ 
			\hline
			\color{red}U35H40T2 & 35 & 40 & 0.27 & 0.26 & 8.70 & 6.79 & 6.80 & 0.01 & 0.10 & 5.1 & 0.25 & 31.7 & 19.5 & 2647 & 40.4 & 0.91 \\ 
			\hline
			\color{red}U35H50T2 & 35 & 50 & 0.31 & 0.24 & 8.70 & 6.79 & 6.80 & 0.01 & 0.10 & 5.1 & 0.25 & 27.9 & 21.3 & 2615 & 32.0 & 0.88 \\ 
			\hline
			\color{red}U35H60T2 & 35 & 60 & 0.36 & 0.21 & 8.70 & 6.79 & 6.80 & 0.01 & 0.10 & 5.1 & 0.25 & 24.0 & 23.6 & 2575 & 24.2 & 0.87 \\ 
			\hline
			\color{red}U35H70T2 & 35 & 70 & 0.44 & 0.19 & 8.70 & 6.79 & 6.80 & 0.01 & 0.10 & 5.1 & 0.25 & 19.8 & 26.9 & 2525 & 17.0 & 0.86 \\ 
			\hline
	\end{tabular}}
\end{table}

\section{Data reduction}
\label{DR}

\begin{figure}[!t]
	\centering
	\includegraphics[width = 1\textwidth]{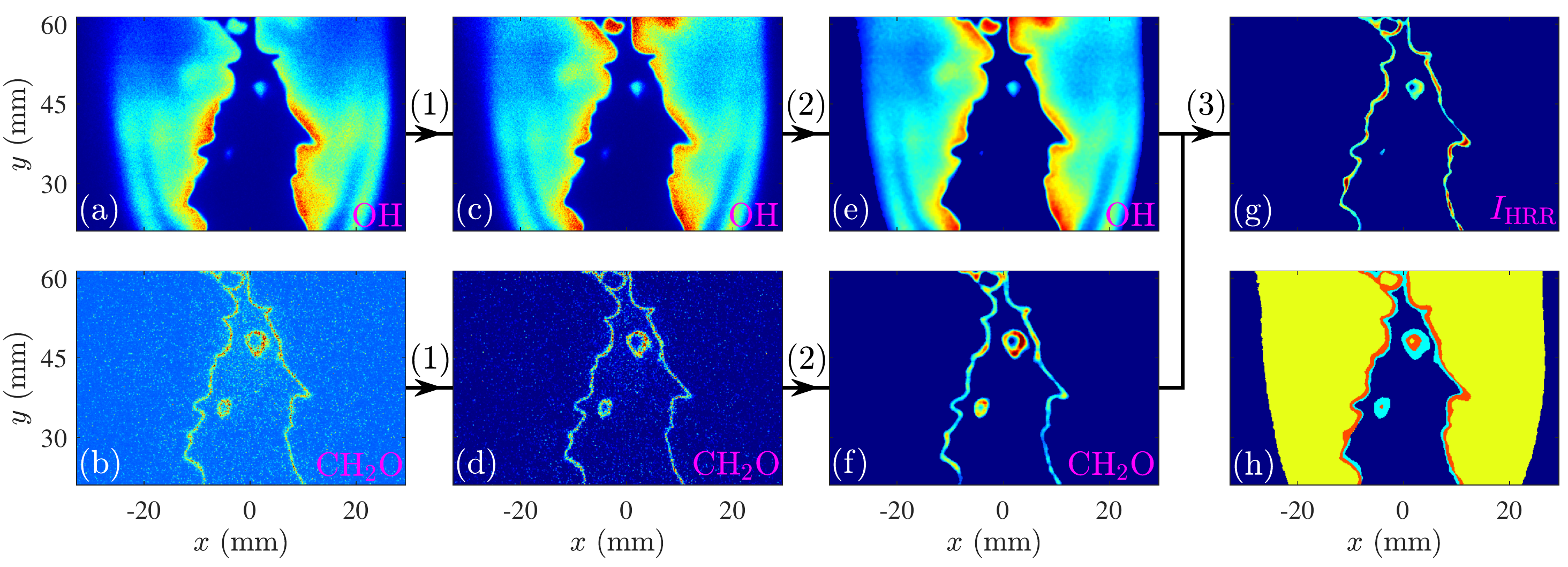} 
	\caption{Procedures for reducing OH (a, c, and e) as well as $\mathrm{CH_2O}$ (b, d, and f) PLIF images. (g) is the estimated heat release rate. (h) presents the preheat (cyan), reaction (red), and combustion products (yellow) zones. Process~(1) corresponds to the background noise, laser profile, and laser pulse energy corrections. Process~2 pertains to applying the median-based filters as well as a global threshold. Process~3 corresponds to estimation of the heat release rate from the processed OH and $\mathrm{CH_2O}$ PLIF images. The results pertain to U5H40T1 condition.}
	\label{fig:DR}
\end{figure}

For each test condition, 500 pairs of OH and $\mathrm{CH_2O}$ PLIF images were collected simultaneously. The PLIF data is used to study both the internal structure as well the burning rate of the tested flames. The processes related to reducing the raw PLIF images is discussed in this section. Figures~\ref{fig:DR}(a) and (b) show two representative frames of the raw OH and $\mathrm{CH_2O}$ PLIF images, respectively. These frames are related to condition~U5H40T1. In order to correct the PLIF images for the effect of the background noise, separate 500 pairs of images were collected for each test condition using the PLIF cameras while the lasers were turned off. These images are averaged and subtracted from the corresponding OH and $\mathrm{CH_2O}$ PLIF data. Then, the obtained images are normalized by the corresponding laser profiles. The profile of the 283~nm laser sheet is obtained from separate acetone PLIF experiments. For these, the ICCD camera $\mathrm{C_1}$ (in Fig.~\ref{fig:Setup}) was equipped with a 305~nm longpass filter. Then, 500 acetone PLIF images were collected, averaged, and used as the laser profile correction of the OH PLIF images. The 355~nm laser sheet profile, which is used to normalized the $\mathrm{CH_2O}$ PLIF images, is acquired similarly except that the longpass filter was removed from the camera and Rayleigh scattering images were collected and used instead, which is similar to our previous study~\cite{mohammadnejad2020thick}. As discussed in Section~\ref{EM}, two photodiodes were used to capture the laser pulse energy variations of both 283 and 355~nm beams. These are used to normalize both the OH and $\mathrm{CH_2O}$ PLIF images (after laser profile correction) in order to remove the effect of the shot-to-shot variation of the lasers pulse energies. Figures~\ref{fig:DR}(c) and (d) show the representative frames of OH and $\mathrm{CH_2O}$ PLIF images corrected for the effects of background noise, laser profile, and shot-to-shot laser energy variations. These corrections are shown as process (1) in Fig.~\ref{fig:DR}. After applying the corrections related to process (1), in order to decrease the remaining background noise, similar to \cite{skiba2018premixed,zhou2017thin,wabel2017measurements,chowdhury2018effectsb,kariuki2015heat,mohammadnejad2020thick,mohammadnejad2019internal}, $7\times 7$ and $11\times 11$ $\mathrm{pixels^2}$ median-based filters are applied to the OH and $\mathrm{CH_2O}$ PLIF images, respectively. Similar results are obtained using the Wiener \cite{coriton2011effect} and Gaussian \cite{rosell2017multi,chaudhuri2010blowoff} filters. Care was taken to ensure the utilized filters do not influence the results. Finally, a global threshold is applied to the images. The application of the median based filters and the global threshold is denoted by process (2) in Fig.~\ref{fig:DR}. The results in Fig.~\ref{fig:DR}(e) and (f) demonstrate the final representative OH and $\mathrm{CH_2O}$ PLIF images after applying both processes (1) and (2).

After generating the reduced OH and $\mathrm{CH_2O}$ PLIF images, the local, normalized, and relative heat release rate (HRR) was obtained, using the model presented in~\cite{mohammadnejad2019internal}. This model suggests that the heat release rate is proportional to $(\mathrm{CH_2O}_\mathrm{PLIF})^\alpha \times (\mathrm{OH_{PLIF}})^\beta$, with $\alpha$ and $\beta$ depending on the tested fuel-air equivalence ratio and the hydrogen-enrichment percentage~\cite{mohammadnejad2019internal}. In the study of \cite{mohammadnejad2019internal}, $\alpha$ and $\beta$ are reported for $\mathrm{H_2\%}$ upto 50\%. Since, in the present study, tests at $\mathrm{H_2\%} =$ 60 and 70\% are also performed, and the corresponding values of $\alpha$ and $\beta$ are not available in \cite{mohammadnejad2019internal}, these parameters are calculated following the procedure discussed in~\cite{mohammadnejad2019internal} and the results are tabulated in Table~\ref{tab:alphabeta}. The results correspond to $\phi=0.7$. Please note that using the conventional method for calculation of HRR (i.e. $\alpha = \beta = 1$~\cite{paul1998planar,skiba2018premixed,temme2015measurements,skiba2015measurements,zhou2017thin,ayoola2006spatially,wabel2017measurements,fayoux2005experimental,mulla2016heat}) leads to similar results and the non unity values of $\alpha$ and $\beta$ do not influence the conclusions of the present study. Figure~\ref{fig:DR}(g) presents the HRR ($I_\mathrm{HRR}$) pertaining to the raw data shown in Figs.~\ref{fig:DR}(a and b).

\begin{table}[h!]
	\caption{Values of $\alpha$ and $\beta$ used for calculation of heat release rate. Results pertain to $\phi=0.7$.}
	\label{tab:alphabeta}
	\centering
	\scalebox{1.0}{
		\begin{tabular}{ccccccccc}
			\hline
			$\mathrm{H_2}\%$ & 0 & 10 & 20 & 30 & 40 & 50 & 60 & 70 \\
			\hline
			$\alpha$ & 0.92 & 0.92 & 0.92 & 0.92 & 0.88 & 0.85 & 0.82 & 0.80 \\ 
			\hline
			$\beta$ & 0.86 & 0.85 & 0.84 & 0.82 & 0.78 & 0.73 & 0.67 & 0.60 \\ 
			\hline
	\end{tabular}}
\end{table}

The PLIF data was used to estimate the preheat and reaction zones thicknesses. Similar to \cite{skiba2018premixed,skiba2015measurements,temme2015measurements,mohammadnejad2020thick}, pixels that feature $I_\mathrm{HRR}$ values larger than 50\% of the global maximum highlight the reaction zone. The leading edge of the preheat zone is where $\mathrm{CH_2O}$ PLIF signal equals 35\% of the maximum $\mathrm{CH_2O}$ PLIF signal~\cite{skiba2015measurements,wabel2017measurements}. The trailing edge of the preheat zone is the leading edge of the reaction zone. The preheat and reaction zones for the representative frames presented in Fig.~\ref{fig:DR} are shown by the cyan and red colors in Fig.~\ref{fig:DR}(h), respectively. After removing the regions pertaining to the reaction zones, the combustion products are defined as the regions where $\mathrm{OH}$ PLIF signal is larger than 15\% of the global maximum. The combustion products are shown by the yellow color in Fig.~\ref{fig:DR}(h). The above procedure for obtaining the preheat, reaction, and combustion products zones are identical to that used in~\cite{mohammadnejad2020thick}. As discussed in~\cite{mohammadnejad2020thick}, five parameters may potentially lead to the uncertainty in estimation of the preheat and reaction zone thicknesses. These parameters are the filtering process, three-dimensional orientation of the flames, imaging resolution, optical blur, and laser sheet thickness. Influence of filtering was discussed earlier. Effect of three-dimensional orientation of the flame is studied in past investigations \cite{hawkes2011estimates,bell2005numerical,bell2007numerical,zhang2015estimation,de2005analysis,rosell2017multi,tyagi2020towards,veynante2010estimation,chatakonda2013fractal}. For example, study of Rosell et al. \cite{rosell2017multi} shows that for a flame with $Ka=60$, the estimated preheat and reaction zone thicknesses can be influenced by three-dimensionality of the flames up to 8\% and 11\%, respectively. Since this Karlovitz number is close to the maximum tested $Ka$ of the present study (which is $Ka=76$), the effect of three-dimensionality on the reported results is not expected to be significant. The effects of imaging resolution, optical blur, and laser sheet thickness on the reported preheat and reaction zone thicknesses have been investigated and reported in our previous study~\cite{mohammadnejad2020thick} using the results, guidelines, and procedures presented in~\cite{lapointe2016fuel,wabel2018assessment}. As discussed in~\cite{mohammadnejad2020thick}, such effects are also not expected to influence the preheat and reaction zone thicknesses reported in the present study.

\section{Results}
\label{Results}

The results are grouped into two subsections. In the first subsection, the preheat and reaction zone thicknesses of the hydrogen-enriched methane-air turbulent premixed flames are presented and discussed. Then, in the second subsection, the turbulent premixed flames burning rate is investigated.

\subsection{Preheat and reaction zones thicknesses}

\begin{figure*}[!t]
	\centering
	\includegraphics[width = 1.0\textwidth]{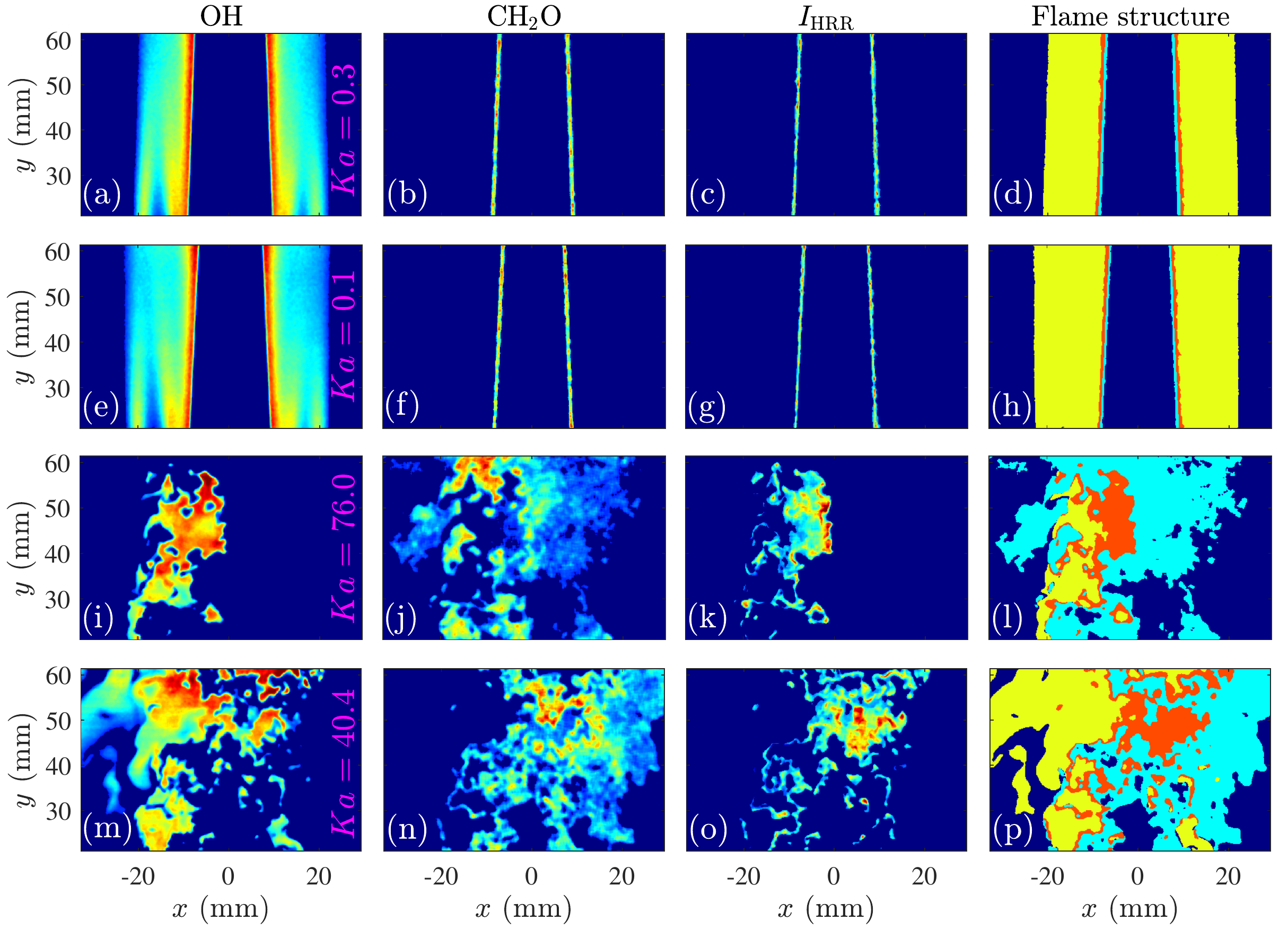} 
	\caption{Representative OH (first column) and $\mathrm{CH_2O}$ (second column) PLIF images as well as the HRR (third column). The preheat, reaction, and combustion products zones in the last column are shown by the cyan, red, and yellow colors, respectively. The first, second, third, and fourth rows pertain to test conditions of~U5H00T0, U5H40T0, U35H00T2, and U35H40T2, respectively.}
	\label{fig:Broadening}
\end{figure*}

Representative planar laser-induced fluorescence images pertaining to the test conditions of U5H00T0 ($Ka = 0.3$), U5H40T0 ($Ka = 0.1$), U35H00T2 ($Ka = 76.0$), and U35H40T2 ($Ka = 40.4$) are presented in the first, second, third, and fourth rows of Fig.~\ref{fig:Broadening}, respectively. The results in the first and second columns are representative hydroxyl and formaldehyde PLIF data. The third column is the heat release rate estimated utilizing the procedure discussed in Section~\ref{DR}. The corresponding preheat, reaction, and combustion products zones were obtained using the procedure discussed in Section~\ref{DR}, and these zones are shown by the cyan, red, and yellow colors in the last column of Fig.~\ref{fig:Broadening}, respectively. For both pure and hydrogen-enriched methane-air flames with $Ka = 0.3$ (U5H00T0) and 0.1 (U5H40T0), both preheat and reaction zones remain relatively thin, see the cyan and red regions in Figs.~\ref{fig:Broadening}(d and h). However, for both pure and hydrogen-enriched flames, as the Karlovitz number increases to large values, the preheat and reaction zones broaden significantly. This observation is consistent with the results presented in \cite{mohammadnejad2020thick,zhou2017thin,zhou2015distributed,zhou2015simultaneous,dunn2007new,dunn2009compositional,dunn2010finite,wang2018direct,wang2017direct,zhou2015visualization,wang2019structure}.

\begin{figure*}[!t]
	\centering
	\includegraphics[width = 1.0\textwidth]{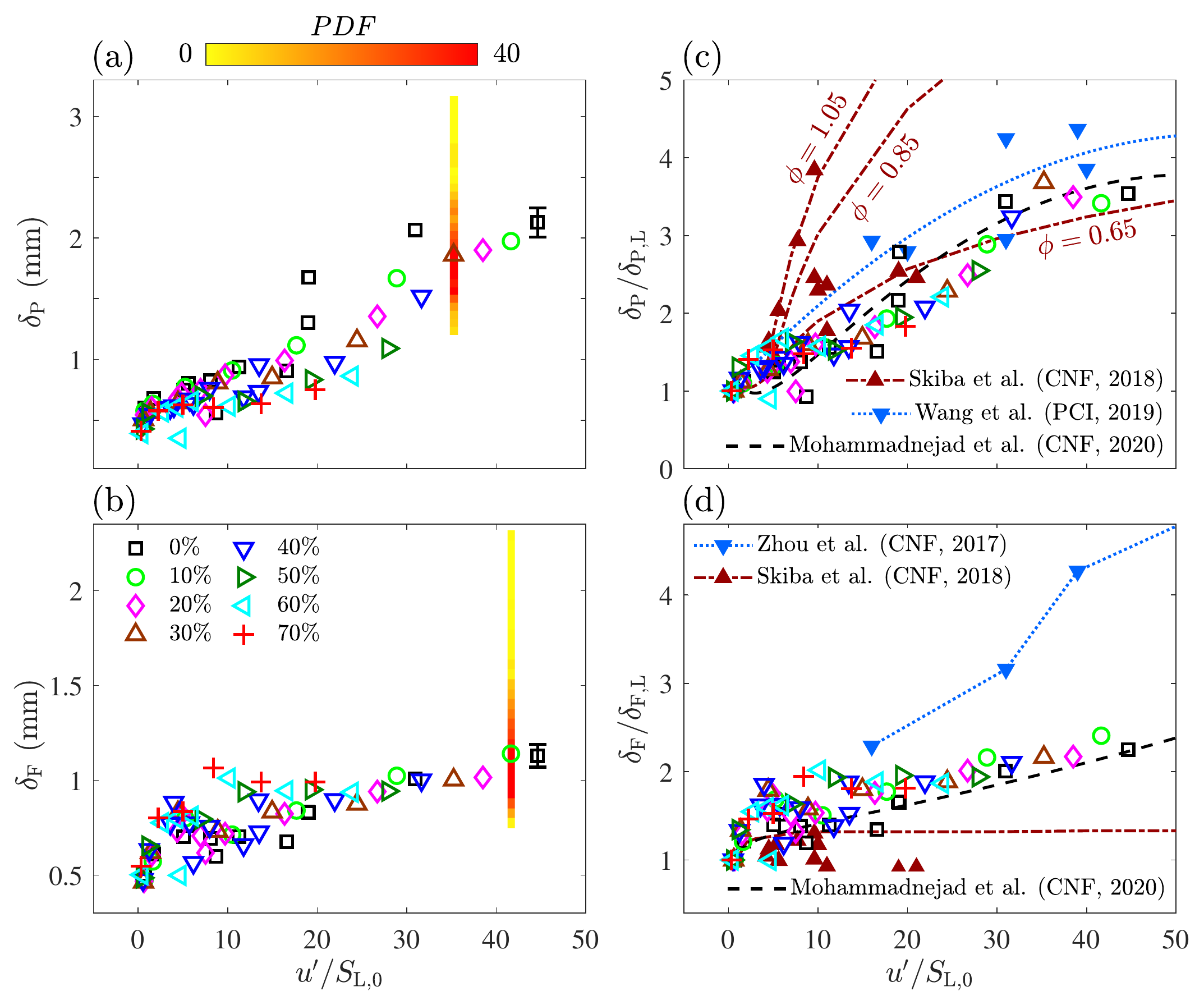} 
	\caption{(a and b) are the preheat and reaction zone thicknesses. (c and d) are the results in (a and b) normalized by $\delta_\mathrm{P,L}$ and $\delta_\mathrm{F,L}$.}
	\label{fig:Thicknesses}
\end{figure*}

Following the procedure elaborated in \cite{mohammadnejad2020thick} and discussed in Section~\ref{DR}, the thicknesses of both preheat and reaction zones were estimated for each frame of the tested conditions. The averaged preheat ($\delta_\mathrm{P}$) and reaction ($\delta_\mathrm{F}$) zone thicknesses are presented in Figs.~\ref{fig:Thicknesses}(a) and (b), respectively. The results in Figs.~\ref{fig:Thicknesses}(a) and (b) normalized by the corresponding laminar flame preheat ($\delta_\mathrm{P,L}$) and reaction ($\delta_\mathrm{F,L}$) zone thicknesses for matching fuel-air equivalence ratio and hydrogen-enrichment percentages are also shown in Figs.~\ref{fig:Thicknesses}(c) and (d), respectively. Values of $\delta_\mathrm{P,L}$ and $\delta_\mathrm{F,L}$ are tabulated in Table~\ref{tab:Lam} and correspond to test conditions with no turbulence generator and at the smallest tested mean bulk flow velocity (5~m/s). The condition associated with the largest tested turbulence intensity (that is U35H00T2) features the maximum uncertainty for estimation of the preheat and reaction zone thicknesses. In order to estimate the maximum uncertainty associated with $\delta_\mathrm{P}$ and $\delta_\mathrm{F}$, the test condition of U35H00T2 was repeated three times, the maximum deviation from the mean value was estimated, and the corresponding error bars were overlaid on Figs.~\ref{fig:Thicknesses}(a) and (b), respectively. The probability density functions of the test conditions with the largest instantaneous preheat and reaction zone thicknesses, which correspond to U35H30T2 and U35H10T2 test conditions, are overlaid by the hot color bar contour on Figs.~\ref{fig:Thicknesses}(a) and (b), respectively.

\begin{table}[h!]
	\caption{Preheat and reaction zone thicknesses at the smallest tested mean bulk flow velocity and with no turbulence generator.}
	\label{tab:Lam}
	\centering
	\scalebox{0.8}{
		\begin{tabular}{ccccccccc}
			
			\hline
			 & U5H00T0 & U5H10T0 & U5H20T0 & U5H30T0 & U5H40T0 & U5H50T0 & U5H60T0 & U5H70T0 \\
			
			\hline
			$\delta_\mathrm{P,L}$ (mm) & 0.60 & 0.58 & 0.54 & 0.50 & 0.47 & 0.43 & 0.39 & 0.41 \\ 
			\hline
			$\delta_\mathrm{F,L}$ (mm) & 0.50 & 0.47 & 0.47 & 0.46 & 0.48 & 0.49 & 0.50 & 0.55 \\ 
			\hline
	\end{tabular}}
\end{table}

Overlaid on Fig.~\ref{fig:Thicknesses}(c) are the normalized preheat zone thickness pertaining to the studies of Skiba et al.~\cite{skiba2018premixed}, Wang et al.~\cite{wang2019structure}, and Mohammadnejad et al.~\cite{mohammadnejad2020thick}, which are shown by the dark red dotted-dashed, blue dotted, and black dashed curves respectively. The studies of~\cite{skiba2018premixed,wang2019structure,mohammadnejad2020thick} correspond to pure methane-air flames. Unlike the present study that the fuel-air equivalence ratio is fixed and equals 0.7, results of~\cite{skiba2018premixed} pertain to the fuel-air equivalence ratios of 0.65, 0.85, and 1.05, and those of~\cite{wang2019structure} pertain to $\phi=0.4$, 0.7, and 1.0. Comparison of our results with those of Skiba et al.~\cite{skiba2018premixed} and Wang et al.~\cite{wang2019structure} at a similar fuel-air equivalence ratio (0.65 and 0.7) suggests that the values of $\delta_\mathrm{P}/\delta_\mathrm{P,L}$ nearly collapse. The results show that, for both pure and hydrogen-enriched methane-air flames, the preheat zone features broadening, and this becomes more pronounced with increasing the turbulence intensity. The broadening of the preheat zone for hydrogen-enriched methane-air turbulent premixed flames has been recently reported in the study of Zhang et al.~\cite{zhang2020effect}. However, the tested conditions in~\cite{zhang2020effect} correspond to to a maximum turbulence intensity of about 12.5. The results of the present study suggest this broadening extends to $u'/S_\mathrm{L,0} \approx 42$ (for 10\% hydrogen-enrichment).

\begin{figure*}[!t]
	\centering
	\includegraphics[width = 0.7\textwidth]{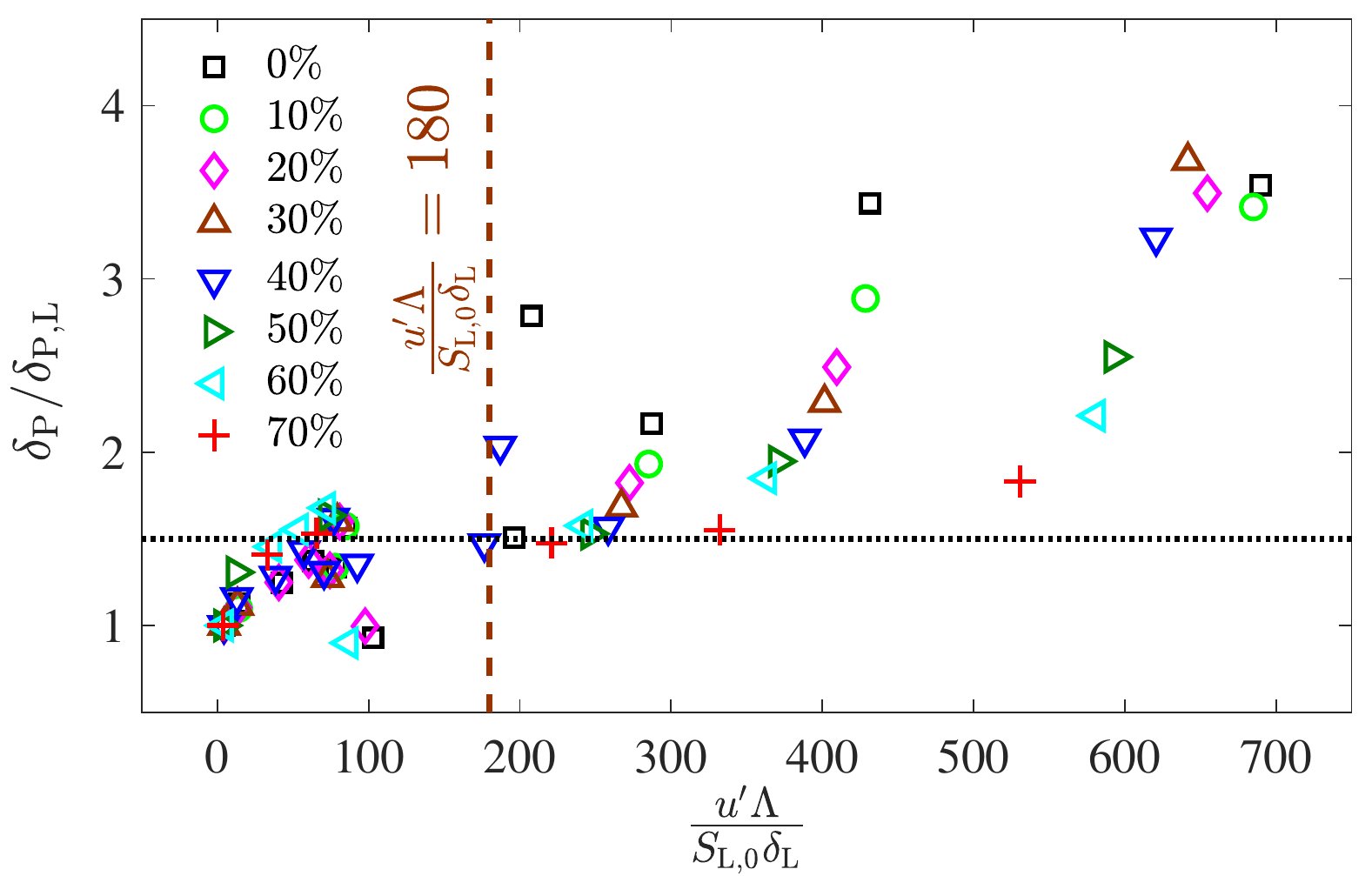} 
	\caption{Variation of the normalized preheat zone thickness versus  ${u'\Lambda}/({S_\mathrm{L,0} \delta_\mathrm{L}})$ for all tested conditions. The dashed line of $u'\Lambda/({S_\mathrm{L,0} \delta_\mathrm{L}}) = 180$ is the border proposed in \cite{skiba2018premixed} and Driscoll et al. \cite{driscoll2020premixed}.}
	\label{fig:PreheatBorghi}
\end{figure*}

Driscoll et al.~\cite{driscoll2020premixed} and Skiba et al.~\cite{skiba2018premixed} suggest the turbulent premixed methane-air flames with $u'\Lambda/({S_\mathrm{L,0} \delta_\mathrm{L}}) \gtrsim 180$ may feature preheat zone broadening. Variation of the normalized preheat zone thickness versus $u'\Lambda/({S_\mathrm{L,0} \delta_\mathrm{L}})$ is presented in Fig.~\ref{fig:PreheatBorghi} for all tested conditions. As can be seen, test conditions with $u'\Lambda/({S_\mathrm{L,0} \delta_\mathrm{L}}) \gtrsim 180$ feature $\delta_\mathrm{P}/\delta_\mathrm{P,L} \gtrsim 1.5$, however, those with $u'\Lambda/({S_\mathrm{L,0} \delta_\mathrm{L}}) \lesssim 180$ feature $\delta_\mathrm{P}/\delta_\mathrm{P,L} \lesssim 1.5$. For pure methane-air turbulent premixed flames, our results agree with those of~\cite{skiba2018premixed,driscoll2020premixed}. The results presented in Fig.~\ref{fig:PreheatBorghi} suggest that the border of $u'\Lambda/({S_\mathrm{L,0} \delta_\mathrm{L}}) \approx 180$ can also be used for predicting the preheat zone broadening of hydrogen-enriched methane-air turbulent premixed flames.

The normalized reaction zone thickness reported in the studies of Skiba et al.~\cite{skiba2018premixed}, Zhou et al.~\cite{zhou2017thin}, and Mohammadnejad et al.~\cite{mohammadnejad2020thick} are overlaid on Fig.~\ref{fig:Thicknesses}(d), which are shown by the dark red dotted-dashed, blue dotted, and black dashed lines respectively. Results of the present study show that increasing the turbulence intensity increases the normalized reaction zone thickness as shown in Fig.~\ref{fig:Thicknesses}(d). This increasing trend is similar for all tested hydrogen-enrichment percentages and agrees well with the results of both Zhou et al.~\cite{zhou2017thin} and Mohammadnejad et al.~\cite{mohammadnejad2020thick}. Compared to the results of the present study and those of~\cite{mohammadnejad2020thick,zhou2017thin}, Skiba et al.~\cite{skiba2018premixed} suggest that the reaction zone thickness remains nearly constant and does not vary by increasing the turbulence intensity. For pure methane-air turbulent premixed flames, Mohammadnejad et al.~\cite{mohammadnejad2020thick} experimentally showed that the reason for the broadening of the preheat zone is due to penetration of turbulent eddies into this zone. Also, Mohammadnejad et al.~\cite{mohammadnejad2020thick} showed that the probability density function of the turbulent eddy size is significantly influenced by the utilized turbulence generating mechanism. As a result, they~\cite{mohammadnejad2020thick} speculated the reason for the controversial observation reported for the reaction zone thickness in their study and in~\cite{zhou2017thin} compared to that in~\cite{skiba2018premixed} is due to the difference between the utilized turbulence generating mechanisms. Nevertheless, the broadening of the reaction zone for hydrogen-enriched methane-air flames shown in Fig.~\ref{fig:Thicknesses}(b and d) at relatively large turbulence intensities is reported in the present study for the first time to the best knowledge of the authors. In essence, the results presented in Fig.~\ref{fig:Thicknesses} show that both the preheat and reaction zones of the pure and hydrogen-enriched methane-air turbulent premixed flames can feature broadening. For example, the preheat and reaction zone thicknesses can feature values that are respectively 6.3 and 4.9 times those of the laminar flame counterparts as shown by the probability density functions in Figs.~\ref{fig:Thicknesses}(a and b).

\subsection{Turbulent premixed flame burning rate}

\begin{figure*}[!t]
	\centering
	\includegraphics[width = 1.0\textwidth]{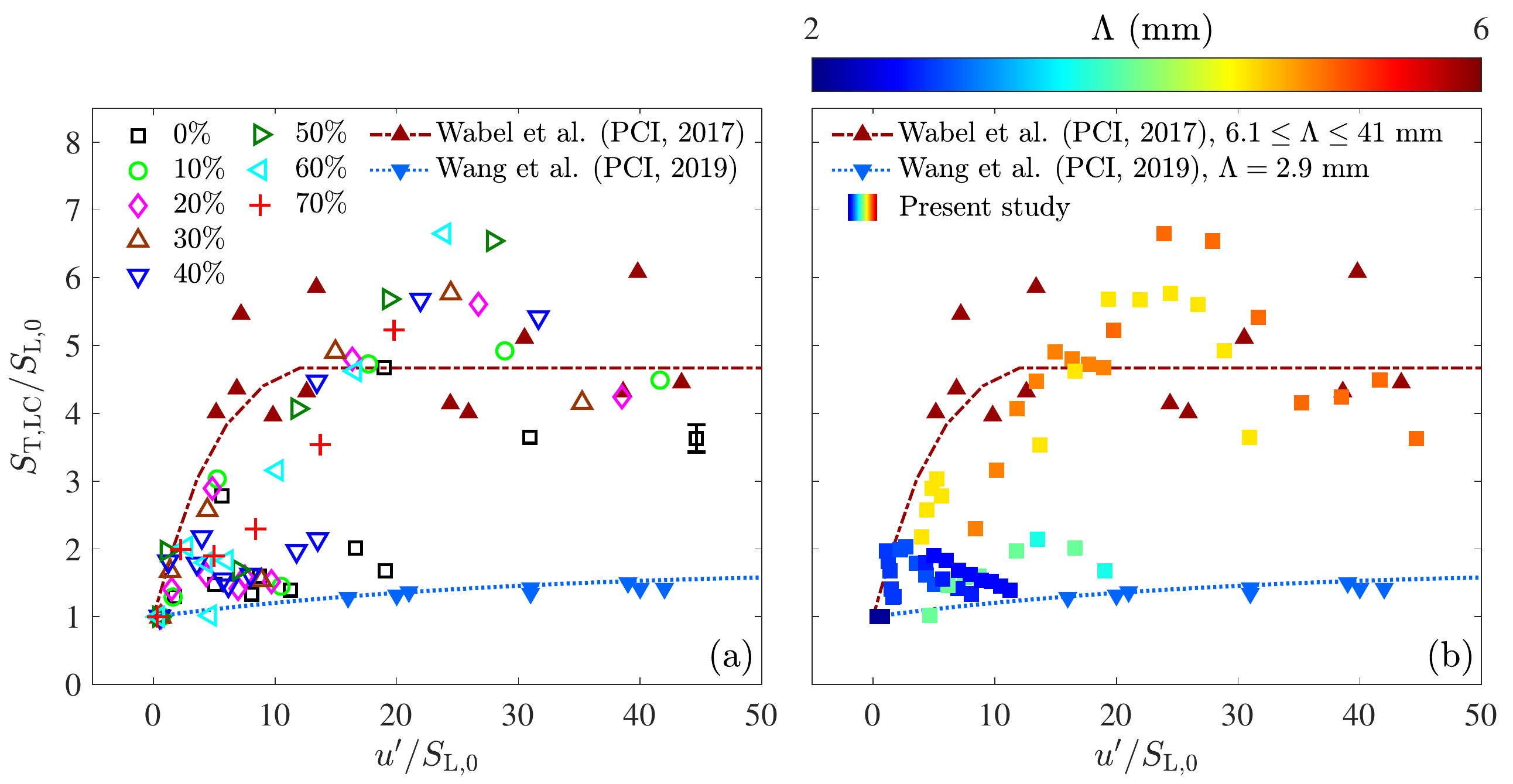} 
	\caption{(a) Mean normalized local consumption speed ($S_\mathrm{T,LC}/S_\mathrm{L,0}$) versus $u'/S_\mathrm{L,0}$. (b) presents the results in (a) color coded based on the corresponding tested integral length scale.}
	\label{fig:BurningLocal}
\end{figure*}

The local consumption speed normalized by the unstretched laminar flame speed is estimated from Eq.~(\ref{Eq:Loc}) and the results are presented in Fig.~\ref{fig:BurningLocal}(a) for all tested conditions. Please note that in Eq.~(\ref{Eq:Loc}), the stretch factor ($I_0$) is assumed to be unity, similar to \cite{wabel2017turbulent}, as the effective Lewis number is close to unity for the tested conditions (please see the last column of Table~\ref{tab:Conditions}). The maximum uncertainty in estimation of the local consumption speed is related to test condition of~U35H00T2, and is shown by the error bar in Fig.~\ref{fig:BurningLocal}(a). The procedure for estimation of the error bar is similar to that discussed in Section~5.1. The results in Fig.~\ref{fig:BurningLocal}(a) are also color coded based on the integral length scale of the test conditions and presented in Fig.~\ref{fig:BurningLocal}(b). Overlaid on both Figs.~\ref{fig:BurningLocal}(a) and (b) are the results extracted from Wabel et al.~\cite{wabel2017turbulent} and Wang et al.~\cite{wang2019structure}, which are shown by the red dotted-dashed and blue dotted lines, respectively. The integral length scale in the study of Wabel et al.~\cite{wabel2017turbulent} varies between 6.1 and 41~mm and that in Wang et al.~\cite{wang2019structure} is 2.9~mm. As can be seen, variation of $S_\mathrm{T,LC}/S_\mathrm{L,0}$ is significantly dependent on the integral length scale and follows two trends. For $\Lambda \lesssim 4$~mm, our results suggest that the normalized local consumption speed plateaus at about 2 following~\cite{wang2019structure}; and, for $\Lambda \gtrsim 4$~mm, $S_\mathrm{T,LC}/S_\mathrm{L,0}$ plateaus at about 5 following~\cite{skiba2018premixed}. Plateau of the normalized local consumption speed with increase of the turbulence intensity (evident in Fig.~\ref{fig:BurningLocal}) has been a matter of discussions in the literature. Several investigations~\cite{wabel2017turbulent,wang2019structure,nivarti2019reconciling,driscoll2020premixed} suggested the reason for this plateau is saturation of the flame surface area with increase of $u'/S_\mathrm{L,0}$ at large values of this parameter. However, the results presented in  Fig.~\ref{fig:Thicknesses} as well as those in~\cite{mohammadnejad2020thick,zhou2017thin,zhou2015distributed,zhou2015simultaneous,dunn2007new,dunn2009compositional,dunn2010finite,wang2018direct,wang2017direct,zhou2015visualization,wang2019structure} suggest that increasing the turbulence intensity broadens both the preheat and reaction zones, and as a result, the flamelet assumption may not hold for such conditions. This means that  Eq.~(\ref{Eq:Loc}), which is developed based on the flamelet assumption, may lead to inaccurate estimation of the local consumption speed. Instead, here, we propose a new formulation that does not rely on the flamelet assumption, which is discussed in the following.

In order to develop a formulation that does not take into account the flamelet assumption, first, similar to~\cite{wabel2017turbulent,wang2019structure,cheng2002premixed,shepherd2001burning}, the OH PLIF images are binarized and the mean progress variable field ($\overline{c}_\mathrm{OH}$) is obtained by averaging the binarized images. Please note this binarization is only performed to calculate the mean progress variable fields, is conducted for flames that are reported to be both relatively thin~\cite{wabel2017turbulent} and thick~\cite{wang2019structure}, and does not necessarily imply a presumed structure for the reaction zone. Then, a curvilinear coordinate system ($\xi-\eta$), with $\xi$ and $\eta$ being respectively tangent and normal to $\overline{c}_\mathrm{OH}$ contours are constructed. The time-average heat release rate is integrated along $\eta-$axis, which is locally normal to the mean progress variable contours of $\overline{c}_\mathrm{OH} = 0.5$. This integral depends on $\xi$ and is given by

\begin{equation}
\label{Eq:HRRBurningLocal}
B_\mathrm{T,\xi}(\xi)={\int_{\eta_\mathrm{min}}^{\eta_\mathrm{max}}\overline{HRR}_\mathrm{T}(\eta,\xi) \mathrm{d}\eta},
\end{equation}
where $\overline{HRR}_\mathrm{T}(\eta,\xi)$ is the mean heat release rate of turbulent flames estimated locally. $\eta_\mathrm{min}$ and $\eta_\mathrm{max}$ are the extents of the $\eta-$axis in the domain of investigation. For a relatively large domain of investigation, the left hand side of Eq.~(\ref{Eq:HRRBurningLocal}) does not change by extending the bounds of integration. The turbulent flame burning rate ($B_\mathrm{T}$) is defined as the spatially averaged (along $\xi$) value of $B_\mathrm{T,\xi}(\xi)$ and is given by
\begin{equation}
\label{Eq:HRRBurning1}
B_\mathrm{T}=\frac{1}{\underbrace{\int_{\xi_\mathrm{min}}^{\xi_\mathrm{max}}\mathrm{d}\xi}_{L_{\xi,\mathrm{T}}}}{\int_{\xi_\mathrm{min}}^{\xi_\mathrm{max}}\underbrace{\int_{\eta_\mathrm{min}}^{\eta_\mathrm{max}}\overline{HRR}_\mathrm{T}(\eta,\xi) \mathrm{d}\eta}_{B_\mathrm{T,\xi}(\xi)} \mathrm{d}\xi}.
\end{equation}
In Eq.~(\ref{Eq:HRRBurning1}), $\xi_\mathrm{min}$ and $\xi_\mathrm{max}$ correspond to the boundaries of $\xi-$axis in the domain of investigation. In the denominator of Eq.~(\ref{Eq:HRRBurning1}), $L_{\xi,\mathrm{T}}$ is the length of $\overline{c}_\mathrm{OH} = 0.5$ contour and depends on the tested condition. Similarly, the laminar flame burning rate can be estimated from
\begin{equation}
\label{Eq:HRRBurningL}
B_\mathrm{L}=\frac{1}{L_{\xi,\mathrm{L}}} {\int_{\xi_\mathrm{min}}^{\xi_\mathrm{max}}\int_{\eta_\mathrm{min}}^{\eta_\mathrm{max}}\overline{HRR}_\mathrm{L}(\eta,\xi) \mathrm{d}\eta \mathrm{d}\xi},
\end{equation}
where $\overline{HRR}_\mathrm{L}(\eta,\xi)$ and $L_{\xi,\mathrm{L}}$ are the mean laminar flame heat release rate estimated locally and the laminar flame length estimated using the length of $\overline{c}_\mathrm{OH} = 0.5$. Dividing Eq.~(\ref{Eq:HRRBurning1}) by Eq.~(\ref{Eq:HRRBurningL}) leads to
\begin{equation}
\label{Eq:HRRBurningN}
\frac{B_\mathrm{T}}{B_\mathrm{L}}=\frac{L_{\xi,\mathrm{L}}}{L_{\xi,\mathrm{T}}}\frac{\int_{\xi_\mathrm{min}}^{\xi_\mathrm{max}}\int_{\eta_\mathrm{min}}^{\eta_\mathrm{max}}\overline{HRR}_\mathrm{T}(\eta,\xi) \mathrm{d}\eta \mathrm{d}\xi}{\int_{\xi_\mathrm{min}}^{\xi_\mathrm{max}}\int_{\eta_\mathrm{min}}^{\eta_\mathrm{max}}\overline{HRR}_\mathrm{L}(\eta,\xi) \mathrm{d}\eta \mathrm{d}\xi}.
\end{equation}
It is assumed that the heat release rate is proportional to $I_\mathrm{HRR}$ estimated using the $\mathrm{OH}$ and $\mathrm{CH_2O}$ PLIF signals following the procedures provided in Section~\ref{DR}. Using this assumption, Eq.~(\ref{Eq:HRRBurningN}) can be simplified to

\begin{equation}
\label{Eq:HRRBurning}
\frac{B_\mathrm{T}}{B_\mathrm{L}}=\frac{L_{\xi,\mathrm{L}}}{L_{\xi,\mathrm{T}}}\frac{\int_{\xi_\mathrm{min}}^{\xi_\mathrm{max}}\int_{\eta_\mathrm{min}}^{\eta_\mathrm{max}}\overline{I}_\mathrm{HRR,T}(\eta,\xi) \mathrm{d}\eta \mathrm{d}\xi}{\int_{\xi_\mathrm{min}}^{\xi_\mathrm{max}}\int_{\eta_\mathrm{min}}^{\eta_\mathrm{max}}\overline{I}_\mathrm{HRR,L}(\eta,\xi) \mathrm{d}\eta \mathrm{d}\xi}
\end{equation}
where $\overline{I}_\mathrm{HRR,T}$ and $\overline{I}_\mathrm{HRR,L}$ are the mean $I_\mathrm{HRR}$ corresponding to turbulent and laminar flames, respectively. Equation~(\ref{Eq:HRRBurning}) allows for estimation of the turbulent flame burning rate normalized by that for the laminar counterpart and is independent of the internal flame structure. In fact, in the following, it is shown that for flames whose internal structure follows the flamelet assumption, Eq.~(\ref{Eq:HRRBurning}) leads to values of burning rate that are similar to those obtained utilizing the formulation proposed by Driscoll~\cite{driscoll2008turbulent}, i.e. Eq.~(\ref{Eq:Loc}).

\begin{figure*}[!t]
	\centering
	\includegraphics[width = 1.0\textwidth]{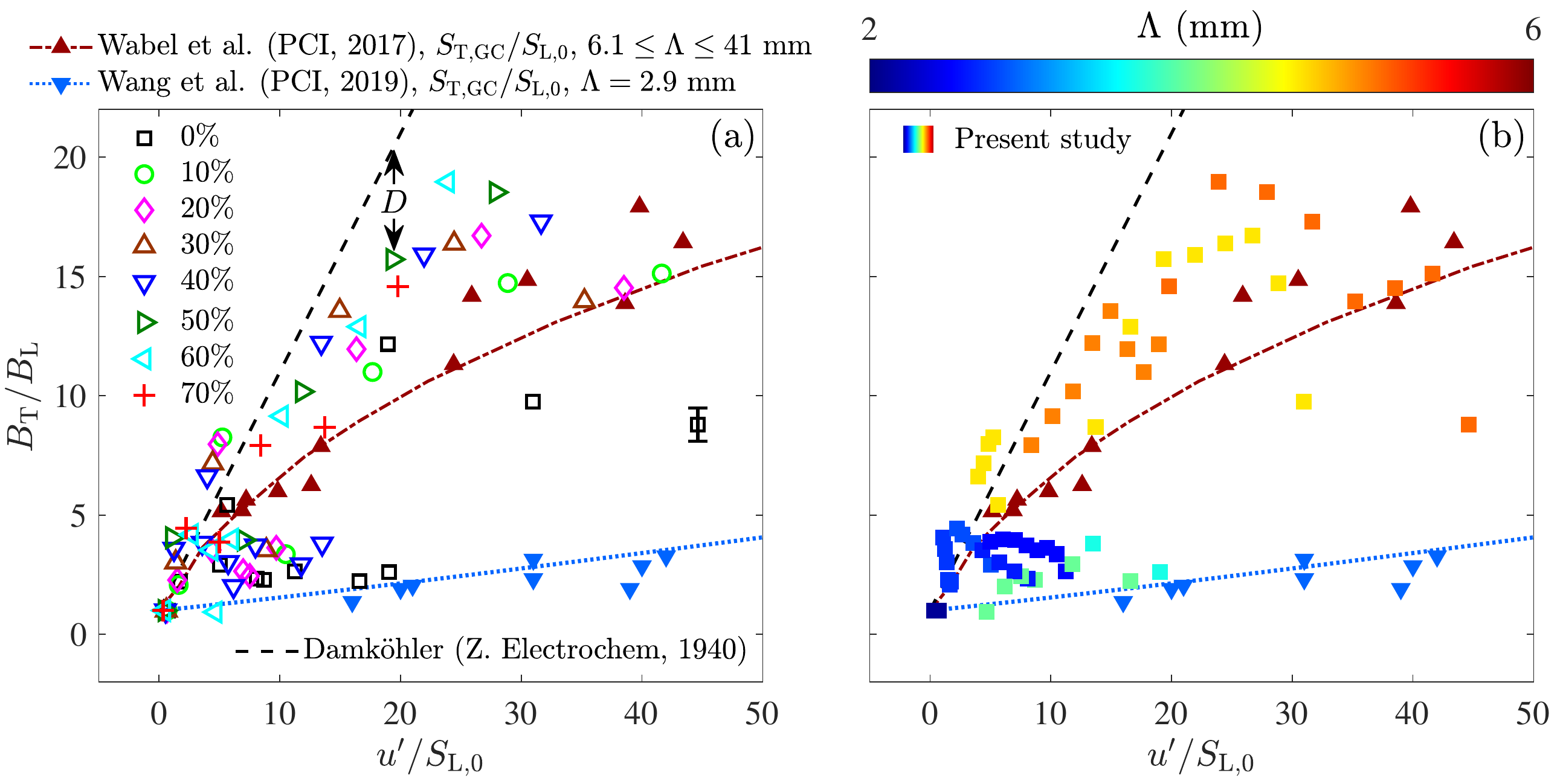} 
	\caption{(a) Variation of the normalized burning rate ($B_\mathrm{T}/B_\mathrm{L}$) versus $u'/S_\mathrm{L,0}$. (b) is the results in (a) color-coded based on the tested integral length scale.}
	\label{fig:BurningGlobal}
\end{figure*}

Variation of $B_\mathrm{T}/B_\mathrm{L}$, estimated from Eq.~(\ref{Eq:HRRBurning}), versus $u'/S_\mathrm{L,0}$ is presented in Fig.~\ref{fig:BurningGlobal}(a). The maximum uncertainty associated with calculation of $B_\mathrm{T}/B_\mathrm{L}$ pertains to test condition of~U35H00T2, which is shown by the error bar in the figure. Among the formulations proposed in the literature, see for example the review paper by Driscoll~\cite{driscoll2008turbulent}, derivation of the global consumption speed formulation does not necessarily depend on the flamelet assumption, and as a result, $S_\mathrm{T,GC}/S_\mathrm{L,0}$ can be compared to $B_\mathrm{T}/B_\mathrm{L}$. Since the field of view of the present investigation is relatively small, the normalized global consumption speed values cannot be estimated for the present investigation; however, $S_\mathrm{T,GC}/S_\mathrm{L,0}$ associated with the studies of~\cite{wabel2017turbulent,wang2019structure} are overlaid on Fig.~\ref{fig:BurningGlobal} by the red dotted-dashed and blue dotted curves. Two trends can be observed for the variation of the normalized burning rate. In order to highlight this, the data presented in Fig.~\ref{fig:BurningGlobal}(a) is color coded based on the tested integral length scale and presented in Fig.~\ref{fig:BurningGlobal}(b). As evident in the figure, for $\Lambda \lesssim 4$~mm, the values of $B_\mathrm{T}/B_\mathrm{L}$ are close to the normalized global consumption speed pertaining to the study of Wang et al.~\cite{wang2019structure}; however, for $\Lambda \gtrsim 4$~mm, the results are closer to those of Wabel et al.~\cite{wabel2017turbulent}. Comparison of the results presented in Fig.~\ref{fig:BurningGlobal} with those in Fig.~\ref{fig:BurningLocal} suggests that the values of $B_\mathrm{T}/B_\mathrm{L}$ and $S_\mathrm{T,GC}/S_\mathrm{L,0}$ are similar, and they are both significantly larger than those of $S_\mathrm{T,LC}/S_\mathrm{L,0}$.

\begin{figure*}[!t]
	\centering
	\includegraphics[width = 0.75\textwidth]{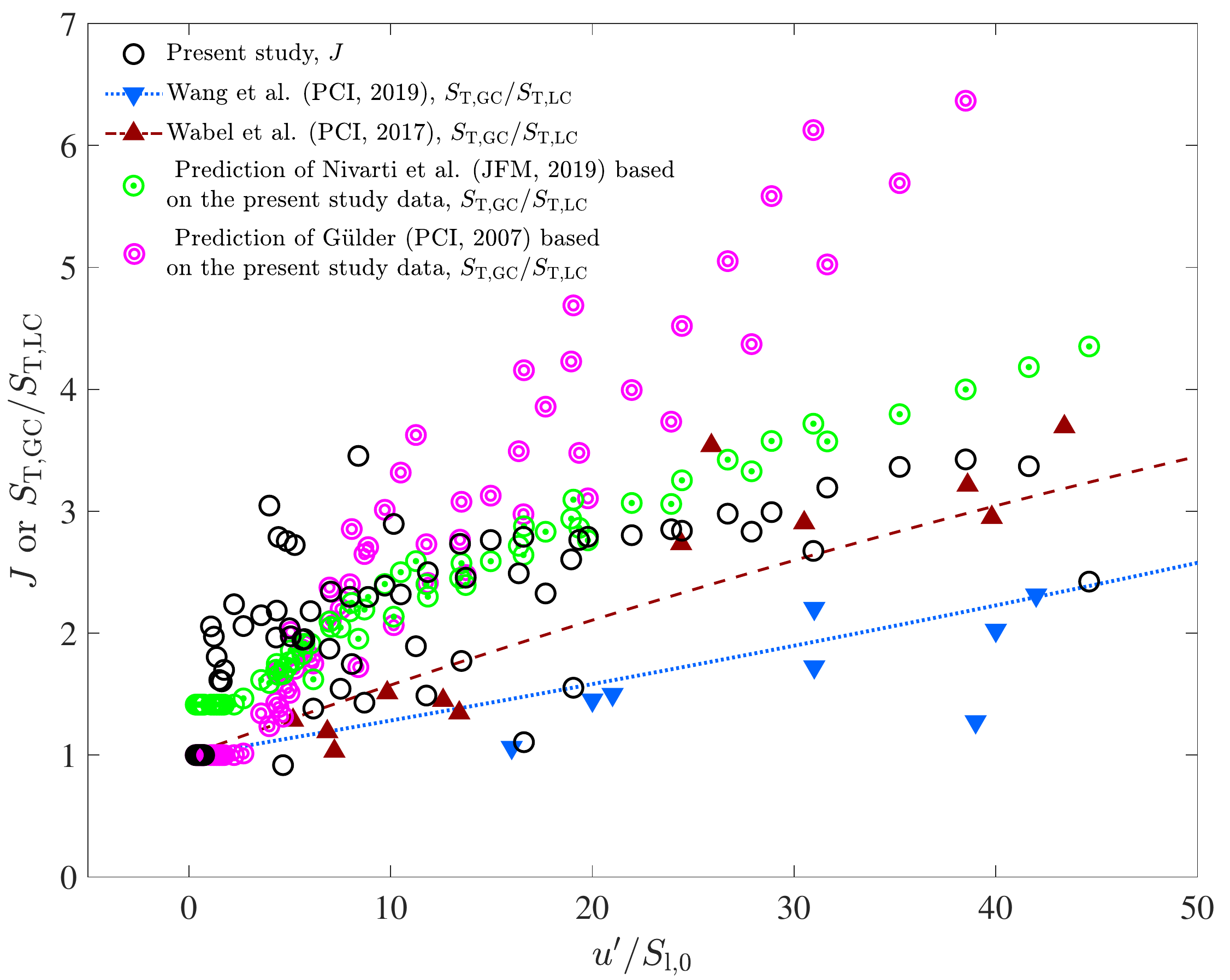} 
	\caption{The normalized burning rate predicted by Eq.~(\ref{Eq:HRRBurning}) (or the normalized global consumption speed) divided by the normalized local consumption speed versus turbulence intensity.}
	\label{fig:Compare}
\end{figure*}

In order to study the reason for the disparity between the values of the normalized burning rate and the normalized local consumption speed, the following parameter is defined.
\begin{equation}
\label{Eq:J}
J=\frac{B_\mathrm{T}/B_\mathrm{L}}{S_\mathrm{T,LC}/S_\mathrm{L,0}}.
\end{equation}
Variation of $J$ versus $u'/S_\mathrm{L,0}$ for the present study is shown in Fig.~\ref{fig:Compare} by the open black circular data symbol. As can be seen, the values of $J$ are larger than unity, and this parameter increases with increasing $u'/S_\mathrm{L,0}$. Variation of the normalized global consumption speed divided by the normalized local consumption speed (i.e. $S_\mathrm{T,GC}/S_\mathrm{T,LC}$) is extracted from the studies of Wabel et al.~\cite{wabel2017turbulent} and Wang et al.~\cite{wang2019structure}, and their results are presented in Fig.~\ref{fig:Compare} by the solid dark red and blue triangular data symbols, respectively. Agreeing, with our results, those of past investigations~\cite{wabel2017turbulent,wang2019structure} also suggest that the local consumption speed underpredicts the turbulent premixed flames burning velocity.

The potential underlying reason for the underprediction of the burning velocity by the local consumption speed has been a matter of discussion over the past decades, see for example~\cite{driscoll2020premixed}. G\"{u}lder~\cite{gulder2007contribution} and Nivarti et al.~\cite{nivarti2019reconciling} suggested that the reason for the observed difference between the local and global consumption speed values can be explained by taking into account the enhanced diffusivity of the reactants due to penetration of small scale eddies into the flame region. Using the mathematical formulations proposed by Nivarti et al~\cite{nivarti2019reconciling} and G\"{u}lder~\cite{gulder2007contribution} as well as the non-dimensional numbers related to test conditions of the present study ($Ka$, $u'/S_\mathrm{L,0}$, and $Re_\lambda = u'l_\mathrm{T}/\nu$), the ratio of the global to local consumption speed was estimated and the results are presented in the Fig.~\ref{fig:Compare} by the green dot-and-circle as well as pink double circle data symbols, respectively. As can be seen, the ratio of the burning rate to local consumption speed in this study ($J$, the black circular data points) is close to that observed in the literature~\cite{wabel2017turbulent,wang2019structure} for the ratio of the global to local consumption speeds ($S_\mathrm{T,GC}/S_\mathrm{T,LC}$). Also, the results estimated based on the mathematical formulation proposed in Nivarti et al.~\cite{nivarti2019reconciling} for  $S_\mathrm{T,GC}/S_\mathrm{T,LC}$ agrees well with $J$ obtained from the present study. However, values of the global consumption speed divided by the local consumption speed obtained based on the  G\"{u}lder's formulation~\cite{gulder2007contribution} deviates from those of the present study and other experimental results~\cite{wabel2017turbulent,wang2019structure} at large turbulence intensities. The reason for the deviation of the results of the present study and the prediction from the formulation provided by G\"{u}lder~\cite{gulder2007contribution} is speculated to be possibly linked to the assumptions made for derivation of the formulation in~\cite{gulder2007contribution}. Specifically, G\"{u}lder~\cite{gulder2007contribution} assumed that the enhanced diffusivity occurs only at the Taylor length scale, while Nivarti et al.~\cite{nivarti2019reconciling} considered enhancement of diffusivity at all length scales smaller than the laminar flame thickness.

\begin{figure*}[!t]
	\centering
	\includegraphics[width = 0.8\textwidth]{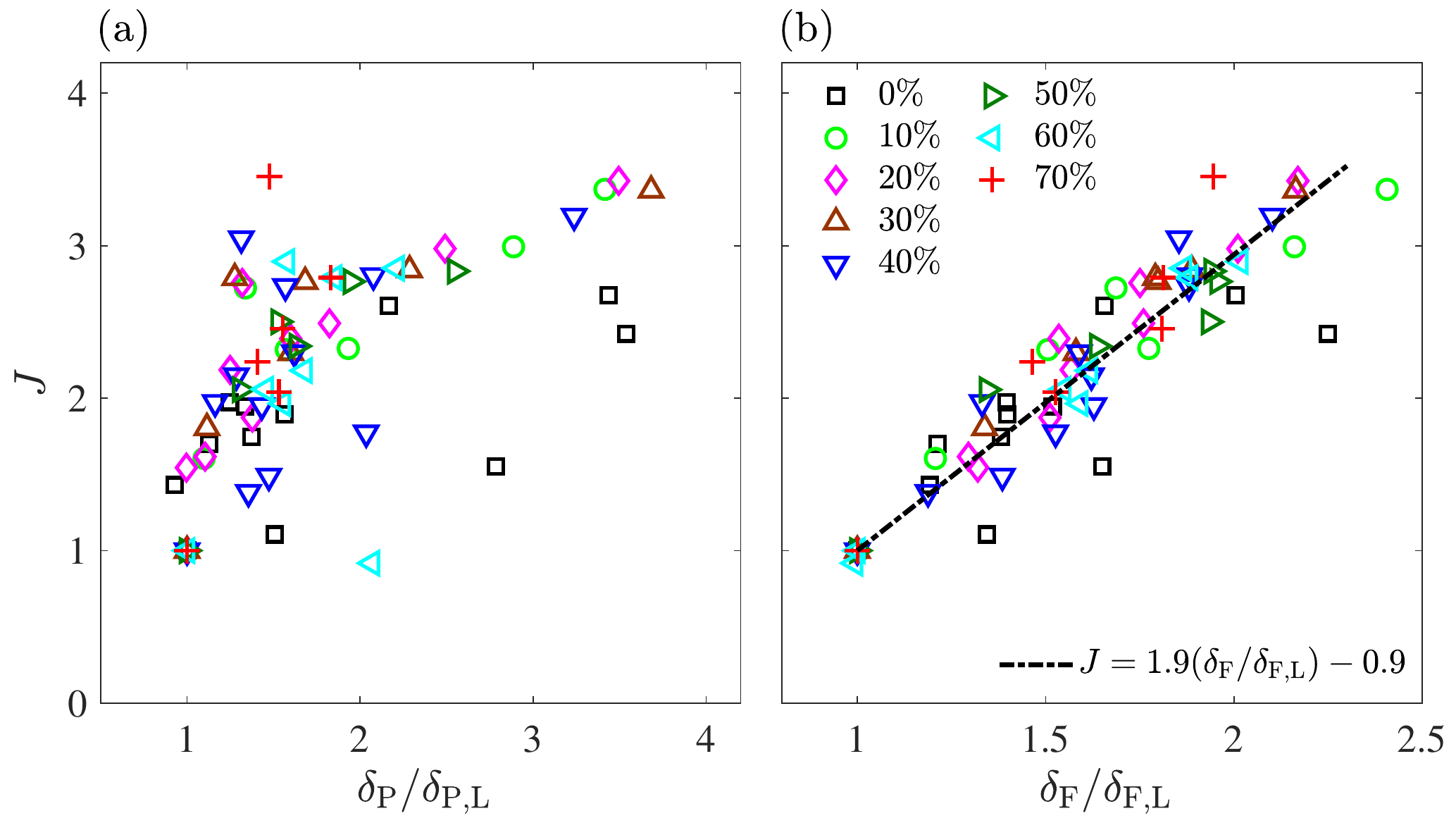} 
	\caption{The ratio of the normalized burning rate and the normalized local consumption speed versus (a) the preheat and (b) the reaction zone thicknesses normalized by those of the corresponding laminar flame counterparts. Overlaid on (b) is the linear fit to the data.}
	\label{fig:Seal}
\end{figure*}

\begin{figure*}[!t]
	\centering
	\includegraphics[width = 0.6\textwidth]{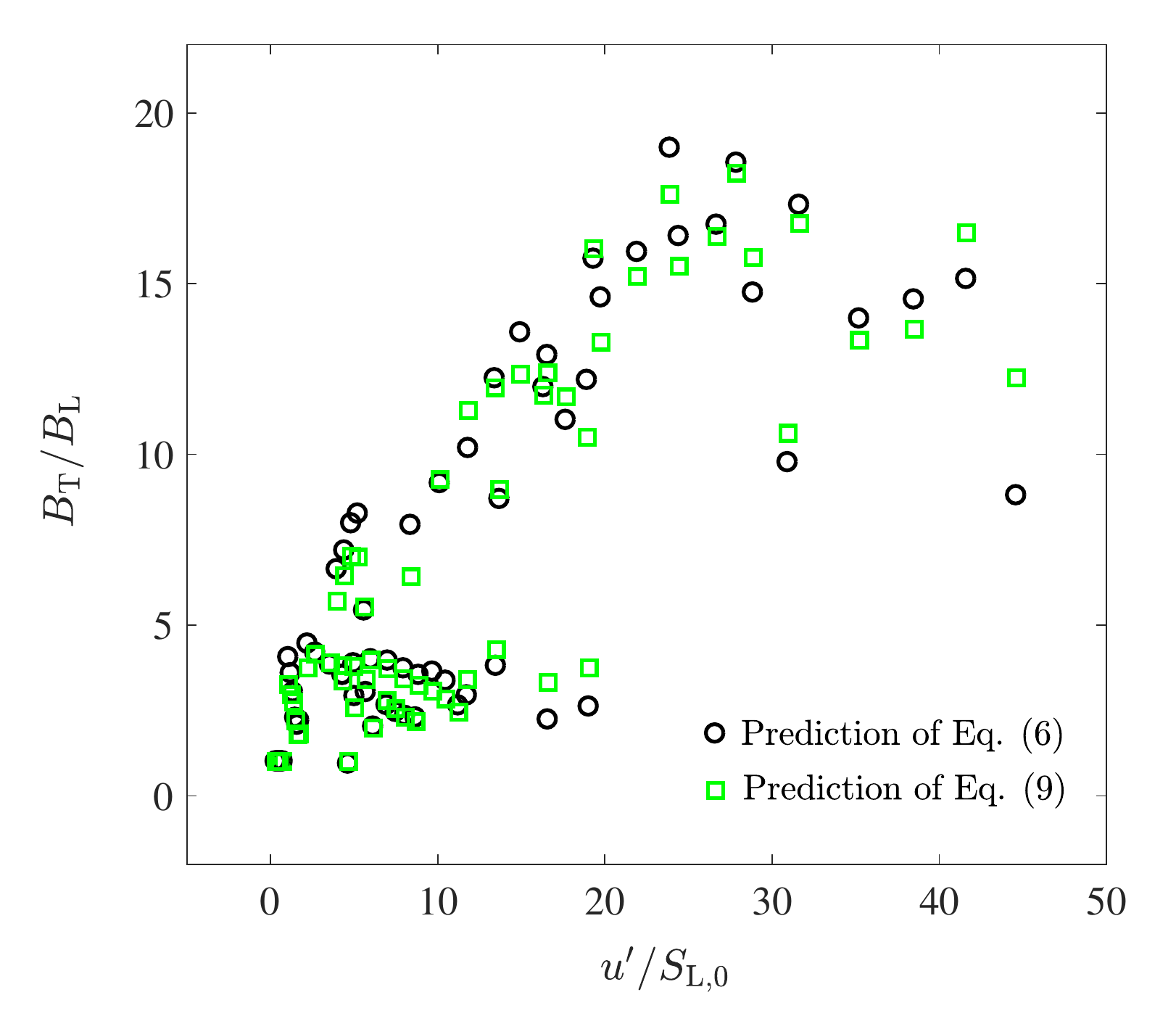} 
	\caption{Comparison of the burning rate values estimated from Eqs.~(\ref{Eq:HRRBurning})~and~(\ref{Eq:ModelThick}).}
	\label{fig:ModelT}
\end{figure*}

The underlying reason for the increasing trend pertaining to variations of $J$ and $S_\mathrm{T,GC}/S_\mathrm{T,LC}$ with $u'/S_\mathrm{L,0}$ is hypothesized to be linked to the internal structure of the tested flames. In order to investigate this hypothesis, variations of $J$ verses the preheat and reaction zone thicknesses are shown in Figs.~\ref{fig:Seal}(a) and (b), respectively. As can be seen, $\delta_\mathrm{P}/\delta_\mathrm{P,L}$ and $\delta_\mathrm{F}/\delta_\mathrm{F,L}$ are positively correlated with $J$. Such correlation is even more pronounced for the reaction zone thickness and follows a linear trend. Thus, a line was fit to the data in Fig.~\ref{fig:Seal}(b) using the least-square technique. The equation of the line is give by
\begin{equation}
\label{Eq:JThickening}
J=1.9(\delta_\mathrm{F}/\delta_\mathrm{F,L})-0.9.
\end{equation}
Equation~(\ref{Eq:JThickening}) suggests that, at the limit of $\delta_\mathrm{F} = \delta_\mathrm{F,L}$, where the flamelet assumption holds, the normalized burning rate estimated from Eq.~(\ref{Eq:HRRBurning}) equals the normalized consumption speed estimated from Eq.~(\ref{Eq:Loc}). However, increasing the reaction zone thickness leads to larger values of the normalized burning rate compared to the normalized local consumption speed. This means that thickening of the turbulent premixed flames is linked to the deviation of the normalized burning rate (and the normalized global consumption speed) from the corresponding normalized local consumption speed values. In fact, the reason for this deviation being linked to the internal flame structure was speculated by Wabel et al.~\cite{wabel2017turbulent} earlier. However, compared to Wabel et al.~\cite{wabel2017turbulent} who suggested the reason for the deviation between $S_\mathrm{T,GC}$ and $S_\mathrm{T,LC}$ may be due to the broadening of the preheat zone, our results show such deviation is linked to broadening of both the preheat and reaction zones. It is experimentally shown that~\cite{mohammadnejad2020thick,skiba2019influence}, with increasing the turbulence intensity, the eddies can penetrate into the preheat zone. This is expected to increase the gas turbulent diffusivity, which increases the burning rate~\cite{damkohler1940einfluss,driscoll2020premixed,nivarti2019reconciling,gulder2007contribution}, and as a result, the larger than unity values of $J$ presented in Fig.~\ref{fig:Seal}(a). It is speculated that, similar to the preheat zone, the turbulent eddies may also penetrate into the reaction zone, leading to increase of $\delta_\mathrm{F}$ (as evident in Fig.~\ref{fig:Thicknesses}(b) as well as the results presented in~\cite{mohammadnejad2020thick,wang2019structure,zhou2017thin}), increasing the turbulent diffusivity, and as a result increasing the flame burning rate. However, the above speculation (that is penetration of turbulent eddies into the reaction zone) remains to be investigated experimentally. It is proposed that Eqs.~(\ref{Eq:J})~and~(\ref{Eq:JThickening}) can be combined with Eq.~(\ref{Eq:Loc}) in order to modify the formulation of the local consumption speed suggested by Driscoll et al.~\cite{driscoll2008turbulent} accounting for the reaction zone thickening. This equation is given by
\begin{equation}
	\label{Eq:ModelThick}
	\frac{B_\mathrm{T}}{B_\mathrm{L}}=\overbrace{[1.9(\delta_\mathrm{F}/\delta_\mathrm{F,L})-0.9]}^{J}I_0\frac{\int_{-\infty}^{\infty} \int_{-\infty}^{\infty}\Sigma(\eta,\xi) d\eta d\xi}{L_\mathrm{\xi}} .
\end{equation}
The values of the normalized burning rate estimated from Eq.~(\ref{Eq:HRRBurning}) along with the prediction of Eq.~(\ref{Eq:ModelThick}) are presented by circular and square data symbols in Fig.~\ref{fig:ModelT} for all test conditions. The results in this figure suggest that the local consumption speed corrected for the effect of flame thickening (Eq.~(\ref{Eq:ModelThick})) follows the values of the burning rate estimated from Eq.~(\ref{Eq:HRRBurning}). This means that the reason for the deviation between the global (here burning rate) and local consumption speeds is the thickening of the reaction zone. This is not shown in the past investigations to our best of knowledge and is demonstrated here.


\begin{figure*}[!t]
	\centering
	\includegraphics[width = 0.9\textwidth]{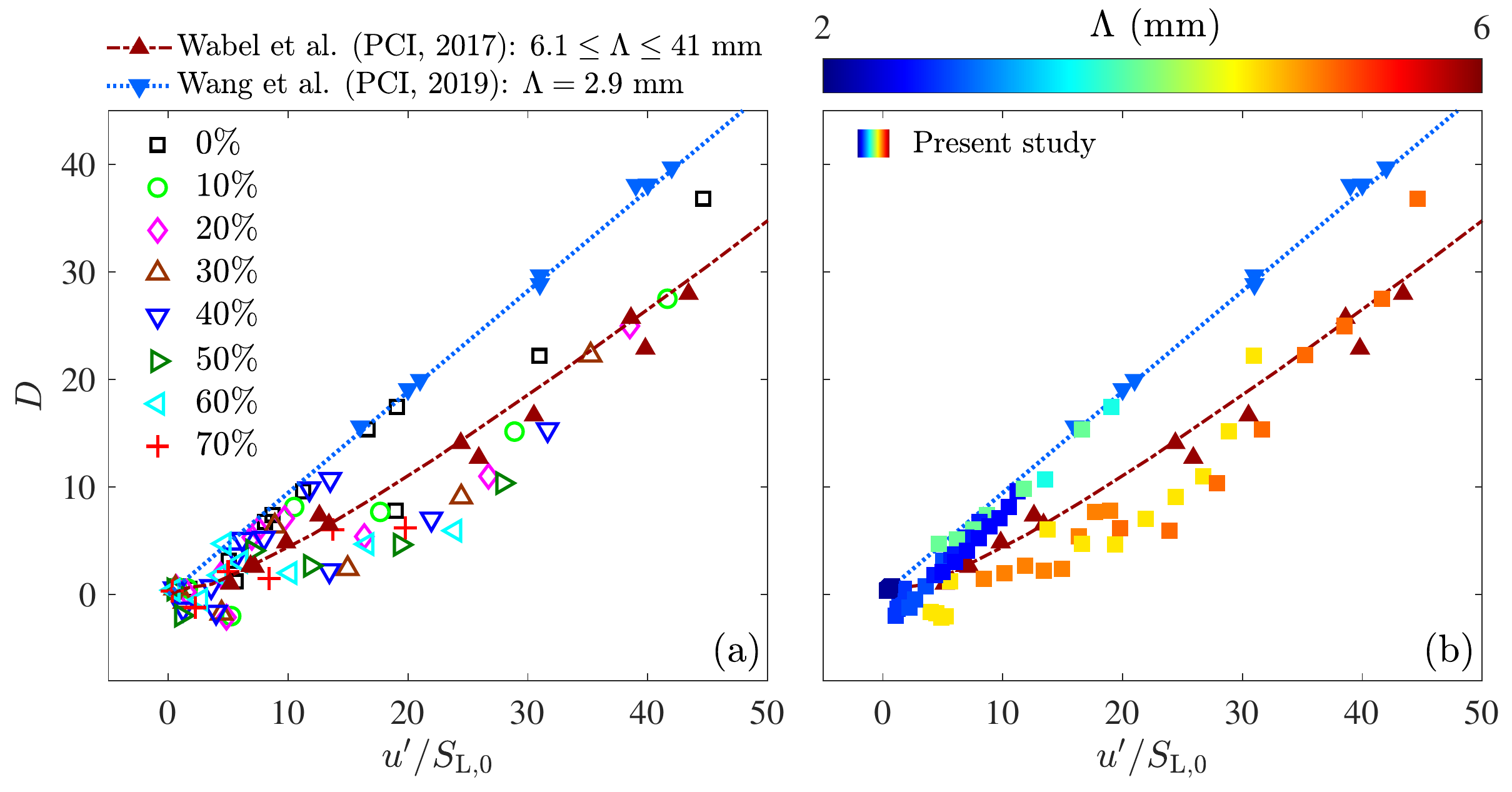} 
	\caption{The difference between the normalized burning rate estimated based on the prediction of Damk\"{o}hler's first hypothesis and the experimental results of the present study. Also overlaid on the figures are the results from the studies of~\cite{wabel2017turbulent,wang2019structure}.}
	\label{fig:Deviation}
\end{figure*}

Although the above argument suggests the values of normalized burning rate can be reconciled using the reaction zone thickness and the integral of the flame surface density, the reason why the variation of $B_\mathrm{T}/B_\mathrm{L}$ versus $u'/S_\mathrm{L,0}$ bends toward the horizontal axis with increasing the turbulence intensity, see Fig.~\ref{fig:BurningGlobal}, is unclear in the literature, has been a matter of discussions~\cite{driscoll2020premixed}, and is investigated here. This behavior is referred to as the bending behavior and is reported in past investigations, see for example~\cite{wabel2017turbulent,wang2019structure,yuen2010dynamics,yuen2013turbulent,kido2002influence,filatyev2005measured}. Here, first, the bending behavior is characterized by estimating the difference between the values of the normalized burning rate and the prediction of the Damk{\"o}hler's first hypothesis~\cite{damkohler1940einfluss} (shown by the dashed line in Fig.~\ref{fig:BurningGlobal}). This difference is referred to as $D$ and is estimated by
\begin{equation}
\label{Eq:D}
D=\left(1+\frac{u'}{S_\mathrm{L,0}}\right)-\frac{B_\mathrm{T}}{B_\mathrm{L}}.
\end{equation}
For example, at $u'/S_\mathrm{L,0} = 19.4$ and $\mathrm{H_2}\% = 50$\%, $D$ is about 4.6, which is shown by the double-sided arrow in Fig.~\ref{fig:BurningGlobal}(a). $D$ was estimated for all tested conditions in Table~\ref{tab:Conditions}, and the results are presented in Fig.~\ref{fig:Deviation}(a). Also overlaid on the figure is the difference between the values of the normalized global consumption speed pertaining to the studies of~\cite{wabel2017turbulent,wang2019structure} from the dashed line in Fig.~\ref{fig:BurningGlobal}(a). The estimated values of $D$ are also color coded based on the integral length scale of the tested conditions and shown in Fig.~\ref{fig:Deviation}(b). As can be seen, $D$ is nearly zero for test conditions corresponding to small turbulence intensities, but it increases with increasing $u'/S_\mathrm{L,0}$. Also, the results in Fig.~\ref{fig:Deviation}(b) show that, at a given turbulence intensity, $D$ nearly decreases with increasing the integral length scale.

\begin{figure*}[!t]
	\centering
	\includegraphics[width = 1.0\textwidth]{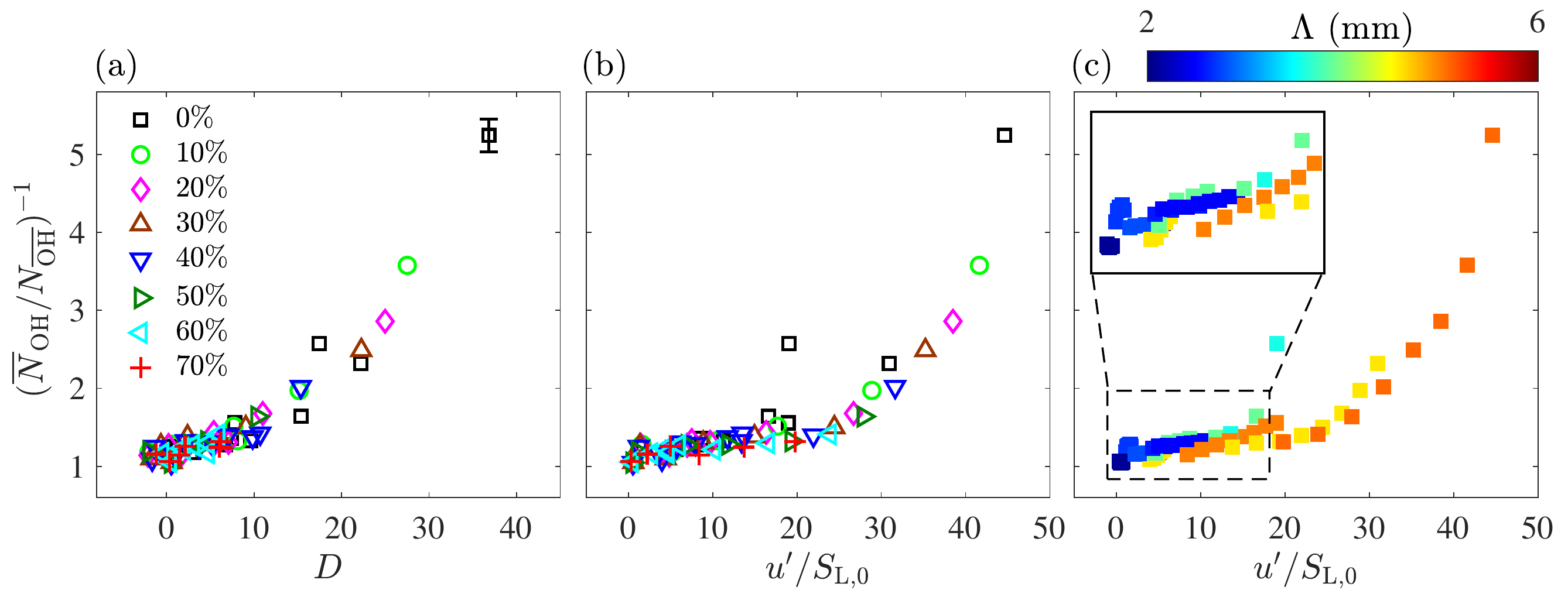} 
	\caption{Variations $(\overline{N}_\mathrm{OH}/N_\mathrm{\overline{OH}})^{-1}$ versus (a) $D$ and (b) $u'/S_\mathrm{L,0}$. The results in (c) are those in (b) color-coded based on the corresponding tested integral length scale values.}
	\label{fig:Escape}
\end{figure*}

The underlying reason for the above observations is hypothesized to be linked to pronounced occurrence of local extinctions and incomplete combustion at relatively intense turbulence conditions. The hydroxyl radical is a post flame species and its concentration is expected to be positively related to completion of the combustion process~\cite{peters2000turbulent,turns1996introduction}. Thus, it is expected that generation of OH radical should be inversely related to incomplete combustion as well as the occurrence of local extinction events. In fact, occurrence of local extinctions has been investigated using the OH PLIF signal in the past, see for example the studies of~\cite{li2010turbulence,garmory2011capturing,jin2019effects,sadanandan2008simultaneous,an2019role}. In the present study, the number of pixels that feature relatively large values of OH PLIF signal (more than 15\% of the maximum) were obtained and averaged for all acquired frames of a given test condition, which is referred to as $\overline{N}_\mathrm{OH}$. In order to facilitate comparison between different test conditions, $\overline{N}_\mathrm{OH}$ was normalized by the number of pixels that are expected to feature relatively large time-averaged OH PLIF signal, which is referred to as $N_\mathrm{\overline{OH}}$. Variations of $(\overline{N}_\mathrm{OH}/N_\mathrm{\overline{OH}})^{-1}$ versus $D$ as well as $u'/S_\mathrm{L,0}$ are presented in Fig.~\ref{fig:Escape}(a), and (b), respectively. The maximum uncertainty for estimation of $(\overline{N}_\mathrm{OH}/N_\mathrm{\overline{OH}})^{-1}$ corresponds to the test condition of U35H00T2, is estimated using the procedure discussed in Section~5.1, and is shown by the error bar in Fig.~\ref{fig:Escape}(a). The results in the figure show that $(\overline{N}_\mathrm{OH}/N_\mathrm{\overline{OH}})^{-1}$ is highly correlated with $D$. This correlation suggests that the occurrence of local extinctions and incomplete combustion are related to the observed bending behavior.

\begin{figure*}[!t]
	\centering
	\includegraphics[width = 0.9\textwidth]{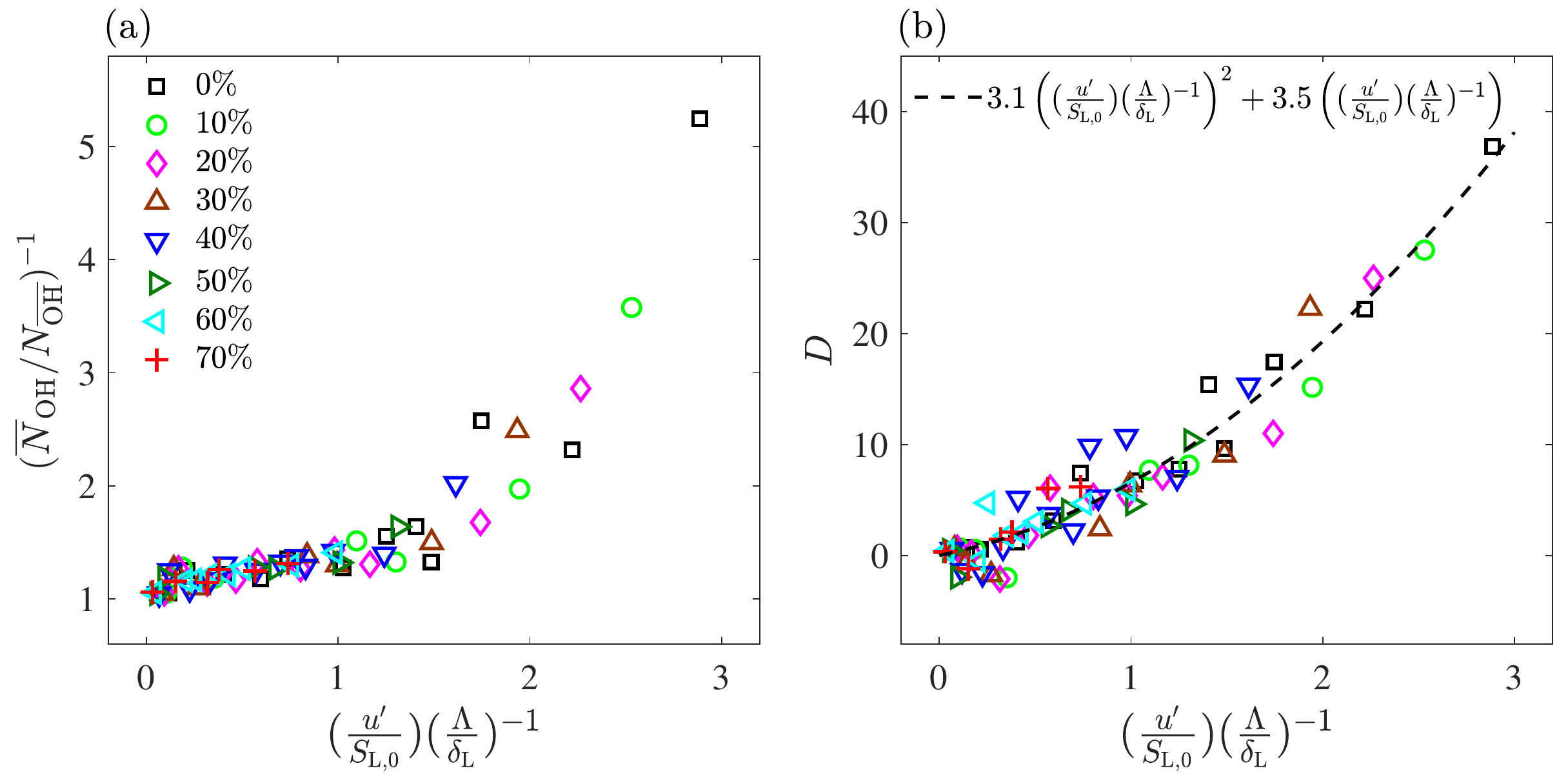} 
	\caption{Variations of (a) $(\overline{N}_\mathrm{OH}/N_\mathrm{\overline{OH}})^{-1}$ and (b) $D$ with $({u'}/{S_\mathrm{L,0}})({\Lambda}/{\delta_\mathrm{L}})^{-1}$.}
	\label{fig:NQ_uL}
\end{figure*}

\begin{figure*}[!t]
	\centering
	\includegraphics[width = 0.7\textwidth]{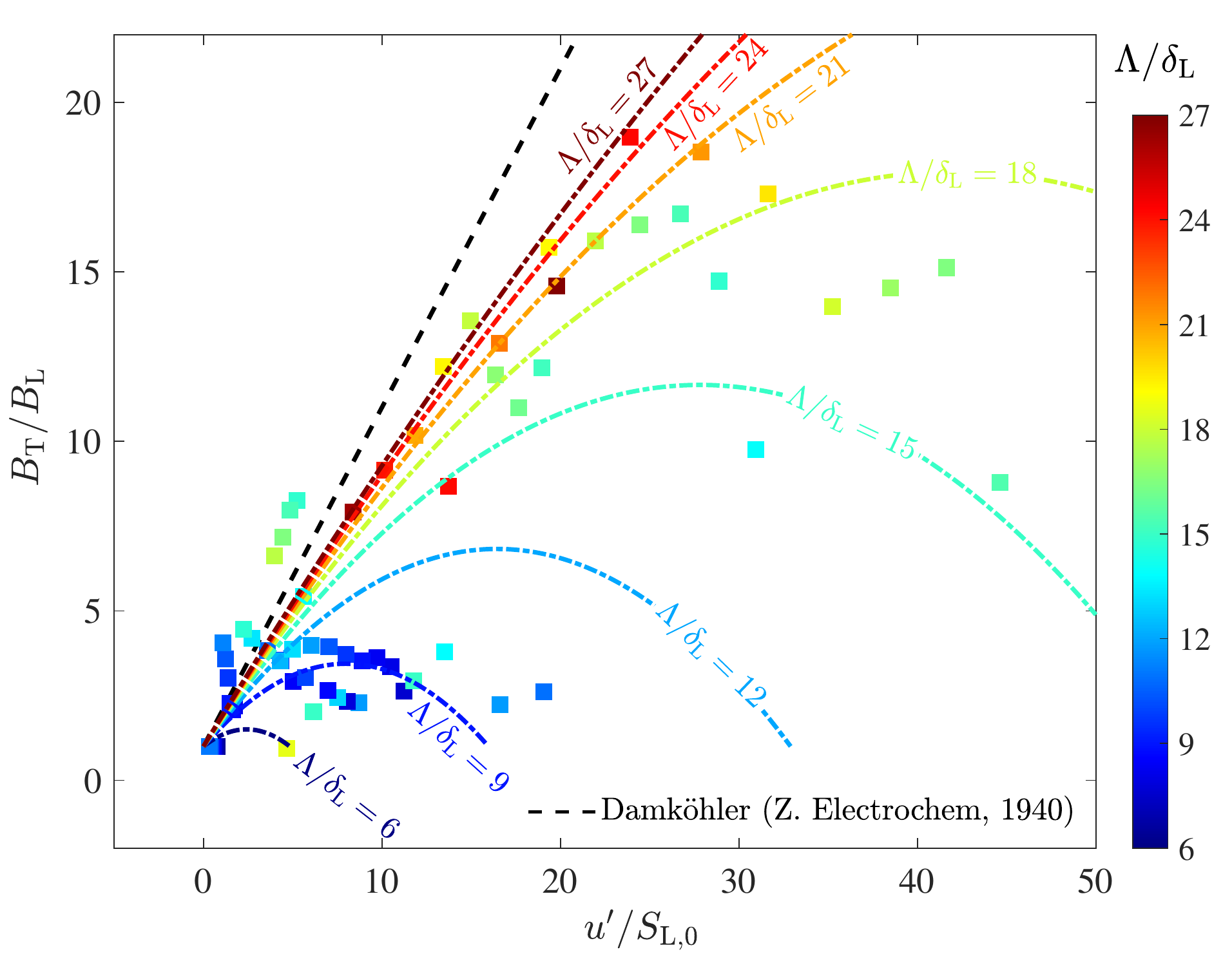} 
	\caption{The normalized burning rate versus turbulence intensity. The dotted-dashed curves pertain to the prediction of Eq.~(\ref{Eq:BT}). The dashed line is the prediction of the Damk\"{o}hler's first hypothesis~\cite{damkohler1940einfluss}.}
	\label{fig:Model}
\end{figure*}

The results presented in Fig.~\ref{fig:Escape}(b) suggest that increasing the turbulence intensity increases $(\overline{N}_\mathrm{OH}/N_\mathrm{\overline{OH}})^{-1}$. As this parameter increases, the possibility of local extinction occurrence increases, the fuel can potentially escape from the flame region, decreasing the generated heat release rate, and as a result the pronounced deviation of the burning rate from the prediction of the Damk\"{o}hler's first hypothesis shown in Fig.~\ref{fig:Deviation}. It is of interest to develop a model that allows for predicting the effect of governing parameters on the occurrence of local extinctions, the bending behavior, and as a result the burning rate. To address these, first, the results presented in Fig.~\ref{fig:Escape}(b) are color-coded based on the corresponding tested integral length scale and presented in Fig.~\ref{fig:Escape}(c). As evident in the inset of the figure, $(\overline{N}_\mathrm{OH}/N_\mathrm{\overline{OH}})^{-1}$ does not change by changing the integral length scale at relatively small values of $u'/S_\mathrm{L,0}$. However, at relatively large values of the turbulence intensity, decreasing the integral length scale increases the values of $(\overline{N}_\mathrm{OH}/N_\mathrm{\overline{OH}})^{-1}$ (compare the values of this parameters at $u'/S_\mathrm{L,0} \approx 20$ and at different integral length scales). At relatively small turbulence intensities, the turbulent eddies cannot penetrate into the flame zone, local extinctions rarely occur, and the values of $(\overline{N}_\mathrm{OH}/N_\mathrm{\overline{OH}})^{-1}$ are relatively small and nearly independent of the tested integral length scales. However, at larger values of $u'/S_\mathrm{L,0}$, eddies feature relatively large rotational kinetic energy and can penetrate into the flame zone~\cite{mohammadnejad2020thick}. At large values of $u'/S_\mathrm{L,0}$, decreasing the integral length scale facilitates penetration of relatively small scale eddies into the flame zone, increasing the occurrence of local extinctions, and as a result the larger values of $(\overline{N}_\mathrm{OH}/N_\mathrm{\overline{OH}})^{-1}$ and $D$ shown in Figs.~\ref{fig:Escape}~and~\ref{fig:Deviation}(b). In essence, the above argument suggests that both $(\overline{N}_\mathrm{OH}/N_\mathrm{\overline{OH}})^{-1}$ and $D$ should be positively (negatively) correlated with $u'/S_\mathrm{L,0}$ ($\Lambda$). To investigate this, variations of $(\overline{N}_\mathrm{OH}/N_\mathrm{\overline{OH}})^{-1}$ and $D$ versus $({u'}/{S_\mathrm{L,0}})({\Lambda}/{\delta_\mathrm{L}})^{-1}$ are presented in Figs.~\ref{fig:NQ_uL}(a) and (b), respectively. The parameter $({u'}/{S_\mathrm{L,0}})({\Lambda}/{\delta_\mathrm{L}})^{-1}$ is similar to that used for calculation of the Karlovitz number, however, different exponents for ${u'}/{S_\mathrm{L,0}}$ and ${\Lambda}/{\delta_\mathrm{L}}$ are used here. This is because, when presented against $({u'}/{S_\mathrm{L,0}})({\Lambda}/{\delta_\mathrm{L}})^{-1}$ (compared to the Karlovitz number), both $(\overline{N}_\mathrm{OH}/N_\mathrm{\overline{OH}})^{-1}$ and $D$ variations collapse. Using the least-square technique, a second order polynomial curve was fit to the variation of $D$ versus $({u'}/{S_\mathrm{L,0}})({\Lambda}/{\delta_\mathrm{L}})^{-1}$, and it is obtained that
\begin{equation}
\label{Eq:DuL}
D \approx 3.1\left((\frac{u'}{S_\mathrm{L,0}})(\frac{\Lambda}{\delta_\mathrm{L}})^{-1}\right)^2+3.5\left((\frac{u'}{S_\mathrm{L,0}})(\frac{\Lambda}{\delta_\mathrm{L}})^{-1}\right).
\end{equation}
Combining Eqs.~(\ref{Eq:D}) and (\ref{Eq:DuL}), it can be shown that the normalized burning rate is obtained from
\begin{equation}
\label{Eq:BT}
\frac{B_\mathrm{T}}{B_\mathrm{L}} \approx \underbrace{\left(1+\frac{u'}{S_\mathrm{L,0}}\right)}_{\mathrm{Term ~I}}\underbrace{-3.1\left((\frac{u'}{S_\mathrm{L,0}})(\frac{\Lambda}{\delta_\mathrm{L}})^{-1}\right)^2-3.5\left((\frac{u'}{S_\mathrm{L,0}})(\frac{\Lambda}{\delta_\mathrm{L}})^{-1}\right)}_{\mathrm{Term~II}}.
\end{equation}
In Eq.~(\ref{Eq:BT}), Term I pertains to the Damk\"{o}hler's first hypothesis, and Term II highlights the negative contribution of bending to the normalized burning rate of turbulent premixed flames. To assess prediction of Eq.~(\ref{Eq:BT}), first, values of the normalized burning rate color-coded based on the normalized integral length scale are presented in Fig.~\ref{fig:Model}. Then, the values of $B_\mathrm{T}/B_\mathrm{L}$ obtained from Eq.~(\ref{Eq:BT}) are overlaid on the figure for several values of $\Lambda/\delta_\mathrm{L}$ by the dotted-dashed curves. The prediction of Damk\"{o}hler~\cite{damkohler1940einfluss} is also shown by the black dashed line. The results in Fig.~\ref{fig:Model} suggest that the formulation proposed in Eq.~(\ref{Eq:BT}) allows for prediction of the normalized burning rate. The results also show that, flames with larger $\Lambda/\delta_\mathrm{L}$ feature less pronounced bending behavior; and, at a fixed value of $u'/S_\mathrm{L,0}$, increasing $\Lambda/\delta_\mathrm{L}$ increases the normalized burning rate. In fact, at the limit of $\Lambda/\delta_\mathrm{L}$ approaching infinity, Eq.~(\ref{Eq:BT}) predicts that $B_\mathrm{T}/B_\mathrm{L}$ approaches the prediction of the Damk\"{o}hler's first hypothesis shown by the dashed line in Fig.~\ref{fig:Model}. Also, for a fixed value of $\Lambda/\delta_\mathrm{L}$, the variation of the normalized burning rate versus $u'/S_\mathrm{L,0}$ features a parabolic behavior. For extremely large values of $u'/S_\mathrm{L,0}$, the effect of Term II in Eq.~(\ref{Eq:BT}) becomes dominant, which is equivalent to pronounced occurrences of local extinctions, leading to decrease of $B_\mathrm{T}/B_\mathrm{L}$. 

The bending behavior and decrease of the burning rate due to local extinctions are consistent with the results of past studies, see for example~\cite{filatyev2005measured,duclos1993comparison,abdel1984turbulent,abdel1987turbulent,bradley1992fast}. The implication of Eq.~(\ref{Eq:BT}) prediction and the results presented in Fig.~\ref{fig:Model} is that the burning rate cannot increase indefinitely with increasing the turbulence intensity. However, at very large normalized integral length scales, the amount of bending is less pronounced. This conclusion is of relevance to DNS studies. Compared to experimental investigations, the DNS studies tend to utilize a relatively small domain of investigation and as a result a relatively small integral length scale, see for example studies of~{\cite{poludnenko2010interaction,poludnenko2011interaction,nivarti2017direct,nivarti2017scalar}. Our results suggest, increasing the size of the domain of investigation and the integral length scale could potentially allow for the DNS investigations to capture relatively larger values of the burning rate.

\section{Conclusions}
\label{Conclusions}

The internal structure and burning rate of extremely turbulent hydrogen-enriched methane-air premixed flames were investigated using simultaneous planar laser-induced fluorescence of $\mathrm{OH}$ and $\mathrm{CH_2O}$ as well as separate stereoscopic particle image velocimetry techniques. In total, 68 test conditions were examined. Fuel-air equivalence ratio was set to 0.7 for all tested conditions, and hydrogen-enrichment percentage was varied from 0\% to 70\% with steps of 10\%. Four mean bulk flow velocities of 5, 15, 25, and 35~m/s were examined. Three turbulence generating mechanisms corresponding to none, one, or two perforated plates were utilized. The turbulence intensity ($u'/S_\mathrm{L,0}$) of the tested conditions varied from 0.3 to 44.6, leading to Reynolds and Karlovitz numbers ranging from 18 to 2729 and 0.1 to 76.0, respectively.

Preheat and reaction zones of hydrogen-enriched methane-air turbulent premixed flames were estimated. It was shown that, at large turbulence intensities, both preheat and reaction zones feature broadening, and their thickness can increase up to 6.3 and 4.9 times those of the laminar flame counterparts, respectively. While broadening of the preheat zone at large turbulence intensities has been reported in the literature for methane-air flames, broadening of this zone for hydrogen-enriched extremely turbulent premixed flames is reported in this study for the first time to the best knowledge of the authors. Similar to extremely turbulent pure methane-air flames, our results show that turbulent premixed hydrogen-enriched methane-air flames also feature broadening of the reaction zone, which is shown here for the first time to our best of knowledge. 

Broadening of the reaction zone observed for both pure and hydrogen-enriched methane-air flames suggests that the flamelet assumption may not lead to an accurate estimation of the burning rate for these flames. A framework that does not utilize the flamelet assumption was developed here and was used to estimated the tested flames burning rate. Also, the values of the local consumption speed were estimated for all tested conditions. The results show that, at relatively small values of the turbulence intensity, the burning rate follows the local consumption speed. However, at larger turbulence intensities, the burning rate follows the global consumption speed reported in the literature. Similar to past investigations, our results show that increasing the turbulence intensity increases the difference between the estimated values of the burning rate and the local consumption speed. Such disparity has been a matter of discussion in the literature. Our results show that the ratio of the normalized burning rate and the normalized local consumption speed is positively correlated with both preheat and reaction zone thicknesses, with the latter correlation being more pronounced. This implies that the flamelet assumption used for calculation of the local consumption speed may have caused the reason for the deviation between the values of global and local consumption speeds reported in the literature. It is shown that, correcting the formulation of the local consumption speed by a non-dimensional factor that takes into account the reaction zone thickening, the burning rate of the tested flames can be reconciled.

Similar to past investigations related to the global consumption speed, the results show that increasing the turbulence intensity does not linearly increase the burning rate, and this parameter features a bending behavior. First, the amount of bending in variation of the normalized burning rate versus the turbulence intensity was quantified. It was hypothesized and shown that the amount of bending is positively related to the occurrence of local extinctions. The results show that both the amount of bending and the occurrence of local extinctions are positively (negatively) related to the turbulence intensity (normalized integral length scale). Using this, a formulation that allows for quantifying the bending behavior was developed. The prediction of the formulation was compared with the experimentally estimated value of the burning rate, and it was shown that both agree well. The proposed formulation suggests that, at a given turbulence intensity, increasing the normalized integral length scale increases the normalized burning rate approaching a limiting value predicted by Damk\"{o}hler's first hypothesis. At a fixed value of the integral length scale, increasing the turbulence intensity, first, increases the burning rate. However, the proposed formulation (in agreement with past experimental results) predicts that further increasing the turbulence intensity decreases the burning rate, which is due to pronounced occurrence of local extinctions.

\section*{Acknowledgments}

The authors would like to thank financial support from the Mitacs Accelerate program, the Gas Turbine Laboratory, and Fortis BC. Sina Kheirkhah and Patrizio Vena thank Professors \"{O}mer L. G\"{u}lder (from the University of Toronto), Matthew Johnson (from the Carleton University), and Dr. Greg Smallwood (from NRC) for lending majority of the setup and the diagnostics.

\bibliography{Ref}

\begin{thebibliography}{100}
\expandafter\ifx\csname url\endcsname\relax
  \def\url#1{\texttt{#1}}\fi
\expandafter\ifx\csname urlprefix\endcsname\relax\def\urlprefix{URL }\fi
\expandafter\ifx\csname href\endcsname\relax
  \def\href#1#2{#2} \def\path#1{#1}\fi

\bibitem{masri2020challenges}
A.~R. Masri, Challenges for turbulent combustion, Proc. Combust. Inst.
  (2020)\href {http://dx.doi.org/https://doi.org/10.1016/j.proci.2020.07.144}
  {\path{doi:https://doi.org/10.1016/j.proci.2020.07.144}}.

\bibitem{driscoll2008turbulent}
J.~F. Driscoll, Turbulent premixed combustion: Flamelet structure and its
  effect on turbulent burning velocities, Prog. Energy Combust. Sci. 34~(1)
  (2008) 91--134.

\bibitem{driscoll2020premixed}
J.~F. Driscoll, J.~H. Chen, A.~W. Skiba, C.~D. Carter, E.~R. Hawkes, H.~Wang,
  Premixed flames subjected to extreme turbulence: {Some} questions and recent
  answers, Prog. Energy Combust. Sci. 76 (2020) 100802.

\bibitem{peters1988laminar}
N.~Peters, Laminar flamelet concepts in turbulent combustion, Proc. Combust.
  Inst. 21~(1) (1988) 1231--1250.

\bibitem{van2016state}
J.~A. van Oijen, A.~Donini, R.~J.~M. Bastiaans, J.~H.~M. ten Thije~Boonkkamp,
  L.~P.~H. de~Goey, State-of-the-art in premixed combustion modeling using
  flamelet generated manifolds, Prog. Energy Combust. Sci. 57 (2016) 30--74.

\bibitem{gicquel2012large}
L.~Y. Gicquel, G.~Staffelbach, T.~Poinsot, Large eddy simulations of gaseous
  flames in gas turbine combustion chambers, Prog. Energy Combust. Sci. 38~(6)
  (2012) 782--817.

\bibitem{duclos1993comparison}
J.~M. Duclos, D.~Veynante, T.~Poinsot, A comparison of flamelet models for
  premixed turbulent combustion, Combust. Flame 95~(1-2) (1993) 101--117.

\bibitem{mohammadnejad2019internal}
S.~Mohammadnejad, P.~Vena, S.~Yun, S.~Kheirkhah, Internal structure of
  hydrogen-enriched methane--air turbulent premixed flames: {F}lamelet and
  non-flamelet behavior, Combust. Flame 208 (2019) 139--157.

\bibitem{skiba2018premixed}
A.~W. Skiba, T.~M. Wabel, C.~D. Carter, S.~D. Hammack, J.~E. Temme, J.~F.
  Driscoll, Premixed flames subjected to extreme levels of turbulence part {I}:
  Flame structure and a new measured regime diagram, Combust. Flame 189 (2018)
  407--432.

\bibitem{skiba2020experimental}
A.~W. Skiba, C.~D. Carter, S.~D. Hammack, J.~F. Driscoll, Experimental
  assessment of the progress variable space structure of premixed flames
  subjected to extreme turbulence, Proc. Combust. Inst. (2020)\href
  {http://dx.doi.org/https://doi.org/10.1016/j.proci.2020.06.129}
  {\path{doi:https://doi.org/10.1016/j.proci.2020.06.129}}.

\bibitem{mohammadnejad2020thick}
S.~Mohammadnejad, Q.~An, P.~Vena, S.~Yun, S.~Kheirkhah, Thick reaction zones in
  non-flamelet turbulent premixed combustion, Combust. Flame 222 (2020)
  285--304.

\bibitem{zhou2017thin}
B.~Zhou, C.~Brackmann, Z.~Wang, Z.~Li, M.~Richter, M.~Ald{\'e}n, X.~S. Bai,
  Thin reaction zone and distributed reaction zone regimes in turbulent
  premixed methane/air flames: {S}calar distributions and correlations,
  Combust. Flame 175 (2017) 220--236.

\bibitem{zhou2015distributed}
B.~Zhou, C.~Brackmann, Q.~Li, Z.~Wang, P.~Petersson, Z.~Li, M.~Ald{\'e}n, X.~S.
  Bai, Distributed reactions in highly turbulent premixed methane/air flames:
  Part {I}. {F}lame structure characterization, Combust. Flame 162~(7) (2015)
  2937--2953.

\bibitem{zhou2015simultaneous}
B.~Zhou, C.~Brackmann, Z.~Li, M.~Ald{\'e}n, X.~S. Bai, Simultaneous
  multi-species and temperature visualization of premixed flames in the
  distributed reaction zone regime, Proc. Combust. Inst. 35~(2) (2015)
  1409--1416.

\bibitem{dunn2007new}
M.~J. Dunn, A.~R. Masri, R.~W. Bilger, A new piloted premixed jet burner to
  study strong finite-rate chemistry effects, Combust. Flame 151~(1-2) (2007)
  46--60.

\bibitem{dunn2009compositional}
M.~J. Dunn, A.~R. Masri, R.~W. Bilger, R.~S. Barlow, G.~H. Wang, The
  compositional structure of highly turbulent piloted premixed flames issuing
  into a hot coflow, Proc. Combust. Inst. 32~(2) (2009) 1779--1786.

\bibitem{dunn2010finite}
M.~J. Dunn, A.~R. Masri, R.~W. Bilger, R.~S. Barlow, Finite rate chemistry
  effects in highly sheared turbulent premixed flames, Flow Turbul. Combust.
  85~(3-4) (2010) 621--648.

\bibitem{wang2018direct}
H.~Wang, E.~R. Hawkes, B.~Savard, J.~H. Chen, Direct numerical simulation of a
  high {Ka} {$\mathrm{CH_4}$/air} stratified premixed jet flame, Combust. Flame
  193 (2018) 229--245.

\bibitem{wang2017direct}
H.~Wang, E.~R. Hawkes, J.~H. Chen, B.~Zhou, Z.~Li, M.~Ald{\'e}n, Direct
  numerical simulations of a high {K}arlovitz number laboratory premixed jet
  flame--an analysis of flame stretch and flame thickening, J. Fluid Mech. 815
  (2017) 511--536.

\bibitem{zhou2015visualization}
B.~Zhou, Q.~Li, Y.~He, P.~Petersson, Z.~Li, M.~Ald{\'e}n, X.-S. Bai,
  Visualization of multi-regime turbulent combustion in swirl-stabilized lean
  premixed flames, Combust. Flame 162~(7) (2015) 2954--2958.

\bibitem{wang2019structure}
Z.~Wang, B.~Zhou, S.~Yu, C.~Brackmann, Z.~Li, M.~Richter, M.~Ald{\'e}n, X.~S.
  Bai, Structure and burning velocity of turbulent premixed methane/air jet
  flames in thin-reaction zone and distributed reaction zone regimes, Proc.
  Combust. Inst. 37~(2) (2019) 2537--2544.

\bibitem{peters2000turbulent}
N.~Peters, Turbulent combustion, Cambridge University Press, 2000.

\bibitem{poinsot2005theoretical}
T.~Poinsot, D.~Veynante, Theoretical and numerical combustion, 2nd ed., R.T.
  Edwards, Philadelphia, USA, 2005.

\bibitem{bell2007numerical}
J.~B. Bell, M.~S. Day, J.~F. Grcar, M.~J. Lijewski, J.~F. Driscoll, S.~A.
  Filatyev, Numerical simulation of a laboratory-scale turbulent slot flame,
  Proc. Combust. Inst. 31~(1) (2007) 1299--1307.

\bibitem{buschmann1996measurement}
A.~Buschmann, F.~Dinkelacker, T.~Sch{\"a}fer, M.~Sch{\"a}fer, J.~Wolfrum,
  Measurement of the instantaneous detailed flame structure in turbulent
  premixed combustion, Proc. Combust. Inst. 26~(1) (1996) 437--445.

\bibitem{filatyev2005measured}
S.~A. Filatyev, J.~F. Driscoll, C.~D. Carter, J.~M. Donbar, Measured properties
  of turbulent premixed flames for model assessment, including burning
  velocities, stretch rates, and surface densities, Combust. Flame 141~(1-2)
  (2005) 1--21.

\bibitem{osborne2016simultaneous}
J.~R. Osborne, S.~A. Ramji, C.~D. Carter, S.~Peltier, S.~Hammack, T.~Lee, A.~M.
  Steinberg, Simultaneous 10 {kHz TPIV, OH PLIF, and {$\mathrm{CH_2O}$} PLIF}
  measurements of turbulent flame structure and dynamics, Exp. Fluids 57~(5)
  (2016) 65.

\bibitem{osborne2017relationship}
J.~R. Osborne, S.~A. Ramji, C.~D. Carter, A.~M. Steinberg, Relationship between
  local reaction rate and flame structure in turbulent premixed flames from
  simultaneous 10 {kHz} {TPIV}, {OH} {PLIF}, and {$\mathrm{CH_2O}$} {PLIF},
  Proc. Combust. Inst. 36~(2) (2017) 1835--1841.

\bibitem{kariuki2015heat}
J.~Kariuki, A.~Dowlut, R.~Yuan, R.~Balachandran, E.~Mastorakos, Heat release
  imaging in turbulent premixed methane--air flames close to blow-off, Proc.
  Combust. Inst. 35~(2) (2015) 1443--1450.

\bibitem{kariuki2016heat}
J.~Kariuki, A.~Dowlut, R.~Balachandran, E.~Mastorakos, Heat release imaging in
  turbulent premixed ethylene-air flames near blow-off, Flow Turbul. Combust.
  96~(4) (2016) 1039--1051.

\bibitem{mansour1998investigation}
M.~S. Mansour, N.~Peters, Y.~C. Chen, Investigation of scalar mixing in the
  thin reaction zones regime using a simultaneous {CH-LIF/Rayleigh} laser
  technique, Proc. Combust. Inst. 27~(1) (1998) 767--773.

\bibitem{poludnenko2010interaction}
A.~Y. Poludnenko, E.~S. Oran, The interaction of high-speed turbulence with
  flames: Global properties and internal flame structure, Combust. Flame
  157~(5) (2010) 995--1011.

\bibitem{poludnenko2011interaction}
A.~Y. Poludnenko, E.~S. Oran, The interaction of high-speed turbulence with
  flames: {T}urbulent flame speed, Combust. Flame 158~(2) (2011) 301--326.

\bibitem{tamadonfar2015experimental}
P.~Tamadonfar, {\"O}.~L. G{\"u}lder, Experimental investigation of the inner
  structure of premixed turbulent methane/air flames in the thin reaction zones
  regime, Combust. Flame 162~(1) (2015) 115--128.

\bibitem{soika1998measurement}
A.~Soika, F.~Dinkelacker, A.~Leipertz, Measurement of the resolved flame
  structure of turbulent premixed flames with constant {R}eynolds number and
  varied stoichiometry, Proc. Combust. Inst. 27~(1) (1998) 785--792.

\bibitem{halter2008investigations}
F.~Halter, C.~Chauveau, I.~G{\"o}kalp, Investigations on the flamelet inner
  structure of turbulent premixed flames, Combust. Sci. Technol. 180~(4) (2008)
  713--728.

\bibitem{aspden2019towards}
A.~Aspden, M.~Day, J.~Bell, Towards the distributed burning regime in turbulent
  premixed flames, J. Fluid Mech. 871 (2019) 1--21.

\bibitem{aspden2011lewis}
A.~J. Aspden, M.~S. Day, J.~B. Bell, Lewis number effects in distributed
  flames, Proc. Combust. Inst. 33~(1) (2011) 1473--1480.

\bibitem{aspden2011turbulence}
A.~J. Aspden, M.~S. Day, J.~B. Bell, Turbulence--flame interactions in lean
  premixed hydrogen: transition to the distributed burning regime, J. Fluid
  Mech. 680 (2011) 287--320.

\bibitem{wabel2017turbulent}
T.~M. Wabel, A.~W. Skiba, J.~F. Driscoll, Turbulent burning velocity
  measurements: Extended to extreme levels of turbulence, Proc. Combust. Inst.
  36~(2) (2017) 1801--1808.

\bibitem{bray1991some}
K.~N.~C. Bray, R.~S. Cant, Some applications of kolmogorov’s turbulence
  research in the field of combustion, Proc. R. Soc. London Ser. A 434~(1890)
  (1991) 217--240.

\bibitem{lapointe2016fuel}
S.~Lapointe, G.~Blanquart, Fuel and chemistry effects in high karlovitz
  premixed turbulent flames, Combust. Flame 167 (2016) 294--307.

\bibitem{nivarti2017scalar}
G.~V. Nivarti, R.~S. Cant, Scalar transport and the validity of
  {D}amk{\"o}hler’s hypotheses for flame propagation in intense turbulence,
  Phys. Fluids 29~(8) (2017) 085107.

\bibitem{nivarti2017direct}
G.~Nivarti, S.~Cant, Direct numerical simulation of the bending effect in
  turbulent premixed flames, Proc. Combust. Inst. 36~(2) (2017) 1903--1910.

\bibitem{nivarti2019reconciling}
G.~Nivarti, R.~Cant, S.~Hochgreb, Reconciling turbulent burning velocity with
  flame surface area in small-scale turbulence, J. Fluid Mech. 858 (2019).

\bibitem{gulder2007contribution}
{\"O}.~L. G{\"u}lder, Contribution of small scale turbulence to burning
  velocity of flamelets in the thin reaction zone regime, Proc. Combust. Inst.
  31~(1) (2007) 1369--1375.

\bibitem{lee2012validation}
D.~Lee, K.~Y. Huh, Validation of analytical expressions for turbulent burning
  velocity in stagnating and freely propagating turbulent premixed flames,
  Combust. Flame 159~(4) (2012) 1576--1591.

\bibitem{peters1999turbulent}
N.~Peters, The turbulent burning velocity for large-scale and small-scale
  turbulence, J. Fluid Mech. 384 (1999) 107--132.

\bibitem{yuen2010dynamics}
F.~T.~C. Yuen, {\"O}.~L. G{\"u}lder, Dynamics of lean-premixed turbulent
  combustion at high turbulence intensities, Combust. Sci. Technol. 182~(4-6)
  (2010) 544--558.

\bibitem{damkohler1940einfluss}
G.~Damk{\"o}hler, Der einfluss der turbulenz auf die flammengeschwindigkeit in
  gasgemischen, Ber. Bunsen. Phys. Chem. 46~(11) (1940) 601--626.

\bibitem{kuo2012fundamentals}
K.~K. Kuo, R.~Acharya, Fundamentals of turbulent and multiphase combustion,
  John Wiley \& Sons, 2012.

\bibitem{yuen2013turbulent}
F.~T.~C. Yuen, {\"O}.~L. G{\"u}lder, Turbulent premixed flame front dynamics
  and implications for limits of flamelet hypothesis, Proc. Combust. Inst.
  34~(1) (2013) 1393--1400.

\bibitem{kido2002influence}
H.~Kido, M.~Nakahara, K.~Nakashima, J.~Hashimoto, Influence of local flame
  displacement velocity on turbulent burning velocity, Proc. Combust. Inst.
  29~(2) (2002) 1855--1861.

\bibitem{zhang2020effect}
W.~Zhang, J.~Wang, W.~Lin, G.~Li, Z.~Hu, M.~Zhang, Z.~Huang, Effect of hydrogen
  enrichment on flame broadening of turbulent premixed flames in thin reaction
  regime, Int. J. Hydrog. Energy (2020)\href
  {http://dx.doi.org/https://doi.org/10.1016/j.ijhydene.2020.09.159}
  {\path{doi:https://doi.org/10.1016/j.ijhydene.2020.09.159}}.

\bibitem{guo2010burning}
H.~Guo, B.~Tayebi, C.~Galizzi, D.~Escudi{\'e}, Burning rates and surface
  characteristics of hydrogen-enriched turbulent lean premixed methane--air
  flames, Int. J. Hydrog. Energy 35~(20) (2010) 11342--11348.

\bibitem{halter2007characterization}
F.~Halter, C.~Chauveau, I.~G{\"o}kalp, Characterization of the effects of
  hydrogen addition in premixed methane/air flames, Int. J. Hydrog. Energy
  32~(13) (2007) 2585--2592.

\bibitem{kheirkhah2014topology}
S.~Kheirkhah, {\"O}.~G{\"u}lder, Topology and brush thickness of turbulent
  premixed {V}-shaped flames, Flow Turbul. Combust. 93~(3) (2014) 439--459.

\bibitem{kheirkhah2016periodic}
S.~Kheirkhah, {\"O}.~L. G{\"u}lder, G.~Maurice, F.~Halter, I.~G{\"o}kalp, On
  periodic behavior of weakly turbulent premixed flame corrugations, Combust.
  Flame 168 (2016) 147--165.

\bibitem{kheirkhah2016experimental}
S.~Kheirkhah, Experimental study of turbulent premixed combustion in {V}-shaped
  flames, Ph.D. thesis, The University of Toronto (2016).

\bibitem{fayoux2005experimental}
A.~Fayoux, K.~Z{\"a}hringer, O.~Gicquel, J.~C. Rolon, Experimental and
  numerical determination of heat release in counterflow premixed laminar
  flames, Proc. Combust. Inst. 30~(1) (2005) 251--257.

\bibitem{ayoola2006spatially}
B.~O. Ayoola, R.~Balachandran, J.~H. Frank, E.~Mastorakos, C.~F. Kaminski,
  Spatially resolved heat release rate measurements in turbulent premixed
  flames, Combust. Flame 144~(1-2) (2006) 1--16.

\bibitem{roder2012simultaneous}
M.~R{\"o}der, T.~Dreier, C.~Schulz, Simultaneous measurement of localized heat
  release with {OH/$\mathrm{CH_2O}$-LIF} imaging and spatially integrated
  $\mathrm{OH}^*$ chemiluminescence in turbulent swirl flames, Appl. Phys. B
  107~(3) (2012) 611--617.

\bibitem{roder2013simultaneous}
M.~R{\"o}der, T.~Dreier, C.~Schulz, Simultaneous measurement of localized
  heat-release with {OH/$\mathrm{CH_2O}$--LIF} imaging and spatially integrated
  $\mathrm{OH}^*$ chemiluminescence in turbulent swirl flames, Proc. Combust.
  Inst. 34~(2) (2013) 3549--3556.

\bibitem{hardalupas2010spatial}
Y.~Hardalupas, C.~Panoutsos, A.~Taylor, Spatial resolution of a
  chemiluminescence sensor for local heat-release rate and equivalence ratio
  measurements in a model gas turbine combustor, Exp. Fluids 49~(4) (2010)
  883--909.

\bibitem{paul1998planar}
P.~H. Paul, H.~N. Najm, Planar laser-induced fluorescence imaging of flame heat
  release rate, Proc. Combust. Inst. 27~(1) (1998) 43--50.

\bibitem{ayoola2009temperature}
B.~Ayoola, G.~Hartung, C.~Armitage, J.~Hult, R.~Cant, C.~Kaminski, Temperature
  response of turbulent premixed flames to inlet velocity oscillations, Exp.
  Fluids 46~(1) (2009) 27.

\bibitem{vena2015heat}
P.~C. Vena, B.~Deschamps, H.~Guo, G.~J. Smallwood, M.~R. Johnson, Heat release
  rate variations in a globally stoichiometric, stratified iso-octane/air
  turbulent {V}-flame, Combust. Flame 162~(4) (2015) 944--959.

\bibitem{harrington1993laser}
J.~E. Harrington, K.~C. Smyth, Laser-induced fluorescence measurements of
  formaldehyde in a methane/air diffusion flame, Chem. Phys. Lett. 202~(3-4)
  (1993) 196--202.

\bibitem{brackmann2003laser}
C.~Brackmann, J.~Nygren, X.~Bai, Z.~Li, H.~Bladh, B.~Axelsson, I.~Denbratt,
  L.~Koopmans, P.~E. Bengtsson, M.~Ald{\'e}n, Laser-induced fluorescence of
  formaldehyde in combustion using third harmonic {Nd:YAG} laser excitation,
  Spectrochim. Acta. A Mol. Biomol. Spectrosc. 59~(14) (2003) 3347--3356.

\bibitem{yamamoto2011local}
K.~Yamamoto, S.~Isii, M.~Ohnishi, Local flame structure and turbulent burning
  velocity by joint {PLIF} imaging, Proc. Combust. Inst. 33~(1) (2011)
  1285--1292.

\bibitem{clemens2002flow}
N.~T. Clemens, Flow imaging, encyclopedia of imaging science and technology,
  John Wiley and Sons Inc., 2002.

\bibitem{wang2004effects}
G.~H. Wang, N.~T. Clemens, Effects of imaging system blur on measurements of
  flow scalars and scalar gradients, Exp. Fluids 37~(2) (2004) 194--205.

\bibitem{cantera}
D.~G. Goodwin, R.~L. Speth, H.~K. Moffat, B.~W. Weber, Cantera: {A}n
  object-oriented software toolkit for chemical kinetics, thermodynamics, and
  transport processes, 2018, \url{https://www.cantera.org}, {V}ersion 2.4.0.

\bibitem{pope2001turbulent}
S.~B. Pope, Turbulent flows, IOP Publishing, 2001.

\bibitem{meyer2007turbulent}
K.~E. Meyer, J.~M. Pedersen, O.~{\"O}zcan, A turbulent jet in crossflow
  analysed with proper orthogonal decomposition, J. Fluid Mech. 583 (2007)
  199--227.

\bibitem{borghi1988turbulent}
R.~Borghi, Turbulent combustion modelling, Prog. Energy Combust. Sci. 14~(4)
  (1988) 245--292.

\bibitem{wabel2017measurements}
T.~M. Wabel, A.~W. Skiba, J.~E. Temme, J.~F. Driscoll, Measurements to
  determine the regimes of premixed flames in extreme turbulence, Proc.
  Combust. Inst. 36~(2) (2017) 1809--1816.

\bibitem{chowdhury2018effectsb}
B.~R. Chowdhury, B.~M. Cetegen, Effects of free stream flow turbulence on
  blowoff characteristics of bluff-body stabilized premixed flames, Combust.
  Flame 190 (2018) 302--316.

\bibitem{coriton2011effect}
B.~Coriton, J.~H. Frank, A.~G. Hsu, M.~D. Smooke, A.~Gomez, Effect of quenching
  of the oxidation layer in highly turbulent counterflow premixed flames, Proc.
  Combust. Inst. 33~(1) (2011) 1647--1654.

\bibitem{rosell2017multi}
J.~Rosell, X.-S. Bai, J.~Sjoholm, B.~Zhou, Z.~Li, Z.~Wang, P.~Pettersson,
  Z.~Li, M.~Richter, M.~Ald{\'e}n, Multi-species {PLIF} study of the structures
  of turbulent premixed methane/air jet flames in the flamelet and
  thin-reaction zones regimes, Combust. Flame 182 (2017) 324--338.

\bibitem{chaudhuri2010blowoff}
S.~Chaudhuri, S.~Kostka, M.~W. Renfro, B.~M. Cetegen, Blowoff dynamics of bluff
  body stabilized turbulent premixed flames, Combust. Flame 157~(4) (2010)
  790--802.

\bibitem{temme2015measurements}
J.~E. Temme, T.~M. Wabel, A.~W. Skiba, J.~F. Driscoll, Measurements of premixed
  turbulent combustion regimes of high reynolds number flames, 53rd AIAA
  Aerospace Sciences Meeting (2015) p. 0168.

\bibitem{skiba2015measurements}
A.~W. Skiba, T.~M. Wabel, J.~E. Temme, J.~F. Driscoll, Measurements to
  determine the regimes of turbulent premixed flames, 51st AIAA/SAE/ASEE Joint
  Propulsion Conference (2015) p. 4089.

\bibitem{mulla2016heat}
I.~A. Mulla, A.~Dowlut, T.~Hussain, Z.~M. Nikolaou, S.~R. Chakravarthy,
  N.~Swaminathan, R.~Balachandran, Heat release rate estimation in laminar
  premixed flames using laser-induced fluorescence of {$\mathrm{CH_2O}$} and
  {H}-atom, Combust. Flame 165 (2016) 373--383.

\bibitem{hawkes2011estimates}
E.~R. Hawkes, R.~Sankaran, J.~H. Chen, Estimates of the three-dimensional flame
  surface density and every term in its transport equation from two-dimensional
  measurements, Proc. Combust. Inst. 33~(1) (2011) 1447--1454.

\bibitem{bell2005numerical}
J.~B. Bell, M.~S. Day, I.~G. Shepherd, M.~R. Johnson, R.~K. Cheng, J.~F. Grcar,
  V.~E. Beckner, M.~J. Lijewski, Numerical simulation of a laboratory-scale
  turbulent {V}-flame, Proc. Natl. Acad. Sci. U.S.A. 102~(29) (2005)
  10006--10011.

\bibitem{zhang2015estimation}
M.~Zhang, J.~Wang, W.~Jin, Z.~Huang, H.~Kobayashi, L.~Ma, Estimation of {3D}
  flame surface density and global fuel consumption rate from {2D} {PLIF}
  images of turbulent premixed flame, Combust. Flame 162~(5) (2015) 2087--2097.

\bibitem{de2005analysis}
L.~P.~H. de~Goey, T.~Plessing, R.~T.~E. Hermanns, N.~Peters, Analysis of the
  flame thickness of turbulent flamelets in the thin reaction zones regime,
  Proc. Combust. Inst. 30~(1) (2005) 859--866.

\bibitem{tyagi2020towards}
A.~Tyagi, J.~O’Connor, Towards a method of estimating out-of-plane effects on
  measurements of turbulent flame dynamics, Combust. Flame 216 (2020) 206--222.

\bibitem{veynante2010estimation}
D.~Veynante, G.~Lodato, P.~Domingo, L.~Vervisch, E.~R. Hawkes, Estimation of
  three-dimensional flame surface densities from planar images in turbulent
  premixed combustion, Exp. Fluids 49~(1) (2010) 267--278.

\bibitem{chatakonda2013fractal}
O.~Chatakonda, E.~R. Hawkes, A.~J. Aspden, A.~R. Kerstein, H.~Kolla, J.~H.
  Chen, On the fractal characteristics of low {D}amk{\"o}hler number flames,
  Combust. Flame 160~(11) (2013) 2422--2433.

\bibitem{wabel2018assessment}
T.~M. Wabel, P.~Zhang, X.~Zhao, H.~Wang, E.~Hawkes, A.~M. Steinberg, Assessment
  of chemical scalars for heat release rate measurement in highly turbulent
  premixed combustion including experimental factors, Combust. Flame 194 (2018)
  485--506.

\bibitem{cheng2002premixed}
R.~K. Cheng, I.~G. Shepherd, B.~Bedat, L.~Talbot, Premixed turbulent flame
  structures in moderate and intense isotropic turbulence, Combust. Sci.
  Technol. 174~(1) (2002) 29--59.

\bibitem{shepherd2001burning}
I.~G. Shepherd, R.~K. Cheng, The burning rate of premixed flames in moderate
  and intense turbulence, Combust. Flame 127~(3) (2001) 2066--2075.

\bibitem{skiba2019influence}
A.~W. Skiba, C.~D. Carter, S.~D. Hammack, J.~D. Miller, J.~R. Gord, J.~F.
  Driscoll, The influence of large eddies on the structure of turbulent
  premixed flames characterized with stereo-{PIV} and multi-species {PLIF} at
  20 {kHz}, Proc. Combust. Inst. 37~(2) (2019) 2477--2484.

\bibitem{turns1996introduction}
S.~R. Turns, An introduction to combustion, McGraw-Hill New York, 1996.

\bibitem{li2010turbulence}
Z.~S. Li, B.~Li, Z.~W. Sun, X.~S. Bai, M.~Ald{\'e}n, Turbulence and combustion
  interaction: {H}igh resolution local flame front structure visualization
  using simultaneous single-shot {PLIF} imaging of {CH}, {OH}, and
  {$\mathrm{CH_2O}$} in a piloted premixed jet flame, Combust. Flame 157~(6)
  (2010) 1087--1096.

\bibitem{garmory2011capturing}
A.~Garmory, E.~Mastorakos, Capturing localised extinction in sandia {F}lame {F}
  with {LES--CMC}, Proc. Combust. Inst. 33~(1) (2011) 1673--1680.

\bibitem{jin2019effects}
W.~Jin, S.~A. Steinmetz, M.~Juddoo, M.~J. Dunn, Z.~Huang, A.~R. Masri, Effects
  of shear inhomogeneities on the structure of turbulent premixed flames,
  Combust. Flame 208 (2019) 63--78.

\bibitem{sadanandan2008simultaneous}
R.~Sadanandan, M.~St{\"o}hr, W.~Meier, Simultaneous {OH-PLIF} and {PIV}
  measurements in a gas turbine model combustor, Appl. Phys. B 90~(3-4) (2008)
  609--618.

\bibitem{an2019role}
Q.~An, A.~M. Steinberg, The role of strain rate, local extinction, and
  hydrodynamic instability on transition between attached and lifted swirl
  flames, Combust. Flame 199 (2019) 267--278.

\bibitem{abdel1984turbulent}
R.~G. Abdel-Gayed, K.~J. Al-Khishali, D.~Bradley, Turbulent burning velocities
  and flame straining in explosions, Proc. R. Soc. A 391~(1801) (1984)
  393--414.

\bibitem{abdel1987turbulent}
R.~G. Abdel-Gayed, D.~Bradley, M.~Lawes, Turbulent burning velocities: {A}
  general correlation in terms of straining rates, Proc. R. Soc. A 414~(1847)
  (1987) 389--413.

\bibitem{bradley1992fast}
D.~Bradley, How fast can we burn?, Proc. Combust. Inst. 24~(1) (1992) 247--262.

\end{thebibliography}
\bibliographystyle{elsarticle-num-PROCI.bst}

\end{document}